\documentclass{vldb}
\usepackage{graphicx}
\usepackage{cite}
\usepackage{times}
\usepackage{amsmath,amssymb,amsfonts}
\usepackage{algorithm,algpseudocode}
\usepackage{graphicx}
\usepackage{textcomp}
\usepackage{xcolor}
\usepackage{microtype}
\usepackage[htt]{hyphenat}
\usepackage{subfigure}
\usepackage{multirow}
\usepackage{algorithm}
\usepackage{color}
\usepackage{listings}
\usepackage{fancyvrb}
\usepackage{enumitem}
\usepackage{layouts}
\usepackage{amsmath}
\usepackage{float}
 \usepackage{booktabs} 
\usepackage{makecell}
\usepackage{subfigure}
\usepackage{lipsum}
\usepackage[T1]{fontenc}
\usepackage[skip=0pt]{caption}
\usepackage[pdftex]{hyperref}
\usepackage{balance}  

\newcommand{\eat}[1]{}

\begin{document}


\title{Pangea: Monolithic Distributed Storage for Data Analytics}



%
%
%
%

\numberofauthors{3} 

\author{
%
%
\alignauthor
Jia Zou\\
       \affaddr{Rice University}\\
       \email{jiazou@rice.edu}
\alignauthor
Arun Iyengar\\
       \affaddr{IBM T. J. Watson Research Center}\\
       \email{aruni@us.ibm.com}
\alignauthor 
Chris Jermaine\\
       \affaddr{Rice University}\\
       \email{cmj4@rice.edu}
}

\maketitle

\begin{abstract}
Storage and memory systems for modern data analytics are heavily
layered, 
managing shared persistent data, cached data, and non-shared execution
data in separate systems such as a distributed file system like HDFS, an in-memory
file system like Alluxio, and a computation framework like Spark. Such
layering introduces significant performance and management costs.
In this paper we propose a
single system called Pangea that can
manage all data---both intermediate and long-lived data, and their
buffer/caching, data placement optimization, and failure recovery---
all in one monolithic distributed storage system, without
any layering. We present a detailed performance evaluation of Pangea and show that its performance compares favorably with several widely used layered systems such as Spark.
\end{abstract}

\lstnewenvironment{code}
  {\lstset{
        aboveskip=5pt,
        belowskip=5pt,
        escapechar=!,
        mathescape=true,
        basicstyle=\linespread{0.94}\ttfamily\footnotesize,
	morekeywords={createSet, sendData, makeLambdaFromMember,
		execute, setInput, setOutput, makeLambda, makeLambdaFromMethod, 
		Handle, Vector, Map, String, pair, LocalitySet,
	PagePtr,
        makeObjectAllocatorBlock, push_back, Object, makeObject,
	addObject, runWork, getSet, addPage, getPageIterators,
	unpinPage, PageIteratorPtr, },	
        showstringspaces=false}
        \vspace{0pt}%
        \noindent\minipage{0.47\textwidth}}
  {\endminipage\vspace{0pt}}

\lstnewenvironment{codesmall}
  {\lstset{
        aboveskip=5pt,
        belowskip=5pt,
        escapechar=!,
        mathescape=true,
        basicstyle=\linespread{0.94}\ttfamily\small,
	morekeywords={createSet, sendData, 
		Handle, Vector, Map, pair,
        push_back, Object, makeObject,
        addObject, getSet, registerClass,
        addPage, getIterators, unpinPage, PageIteratorPtr,
	getPageIterators,
        ObjectIteratorPtr, getObjectIterator, RecordPtr,
	VirtualShuffleBufferPtr,
        getVirtualShuffleBuffer, PartitionComp, PagePtr, next, hash,
        partitionSet, registerReplica, runWork, addData, addBytes,
        VirtualHashBufferPtr, createVirtualHashBuffer, find, insert},	
        showstringspaces=false}
        \vspace{0pt}%
        \noindent\minipage{0.47\textwidth}}
  {\endminipage\vspace{0pt}}
\section {Introduction}
\label{sec:introduction}


One of the defining characteristics of modern analytics systems such as
Spark~\cite{zaharia2012resilient},
Hadoop~\cite{white2012hadoop}, TensorFlow~\cite{abaditensorflow},
Flink~\cite{alexandrov2014stratosphere} and others \cite{clusterSize, crotty2015architecture} is that
they tend
to be heavily layered. For example, consider Spark. It caches job input/output and execution data
(for shuffling and aggregation) in two separate memory
pools: a storage pool
(that is, the RDD cache or DataFrame) and an execution pool. 
These software components, in turn, are implemented on top of the Java
virtual machine (JVM), which constitutes its own software layer. 
Since application output is transient and does
not persist,
such data must be 
written to \emph{yet another} layer, an
external storage system such as
HDFS~\cite{borthakur2008hdfs}, or additionally cached in an external
in-memory system like Alluxio~\cite{li2014tachyon}\textcolor{black}{, Ignite~\cite{ignite}, and so on}.

\vspace{3 pt}
\noindent
\textbf{The Perils of Layering.}
Layering allows for simpler components and facilitates mixing and
matching of different systems that use compatible
interfaces.
However, it also introduces significant
costs in terms of performance.
Some important factors leading to high costs are:

\vspace{3pt}
\noindent
{\bf (1) Interfacing Overhead}. Data needs to be repeatedly de-objectified (serialized), copied,
  and objectified (de-serialized) as it is moved 
  between layers,
  which wastes CPU cycles and memory
  resources. 
  \eat{Even flat, binary data needs to be transferred between processes.}
 For this reason, we have found that an external cache (such as
 Alluxio or Ignite) while important for persisting and sharing data across applications,
  can incur a more than
  50\%
  overhead compared to Spark without such a cache, when processing
  widely used machine learning algorithm such as $k$-means.

\vspace{3pt}
\noindent
{\bf (2) Data Placement Complexity}. When job input data must be
stored at all layers,
redundant copies of the same data object will be maintained at each
layer. Our observation is that in the $k$-means workload, about
$30\%$ of total required memory for processing is used for storing
   redundant input data.

In addition, layering can result in the same data object being copied multiple
times \emph{within} the same layer. For example, at the storage layer, HDFS replicates each
data block for
high availability. At the application layer, each Spark
application might load and partition all of the objects in a dataset
in a different way using different partition sizes and partition
schemes. In a layered system, there is obviously a knowledge gap between storage and
applications. It may make more
sense to bridge the knowledge gap by removing the layers and allow the
storage system to offer multiple partitionings that can be used as replications for failure recovery and also
re-used at the application layer. We observe up to $20\times$ speedup
for TPC-H queries benefiting from such idea while not increasing the
storage size by using the same replication factor.

\vspace{3pt}
\noindent
{\bf (3) Un-Coordinated Resource Utilization}. 
In addition, moving seldom-ac\-cess\-ed, in-RAM data to secondary storage (paging it out) is a fundamental tool for
dealing with large datasets.  
However, when no single system component has a unified view of the system, 
it is difficult to make good paging
decisions: one component may page out data when it needs more RAM, while there is a far
less important piece of data in another component that should have
been paged out. For example, although HDFS can utilize Linux OS's
\texttt{fadvise} and \texttt{madvise} system
calls~\cite{mortonusermode} to specify file access patterns for optimizing paging in OS buffer cache, it has no way
to influence allocation of memory resources at upper layers like
Alluxio\textcolor{black}{, Ignite} and Spark. We find that when size of working set exceeds
available memory, good paging decision can achieve $1.8\times$ to
$5\times$ speedup compared with LRU, MRU, and DBMIN~\cite{chou1986evaluation} policies for
the $k$-means workload.

\vspace{3pt}
\noindent
\textbf{Pangea: A Monolithic Distributed Storage Manager.}
{\color{black}
The benefits of layering---including more flexible, and interchangeable tools and components---will
outweigh the costs for many applications.  For example, if a distributed file system does not have the right recovery
model for a given application, one can (in theory) swap in a new file system that addresses the problem.
However, for applications that demand high performance, 
layering's performance cost compared to a monolithic architecture may be unacceptable.

Monolithic architectures are not new, and
the perils of heavily modular systems are well-understood; indeed, performance vs. flexibility
concerns underlie the classical
microkernel vs. macrokernel operating system debate \cite{pai2000io,liedtke1996toward,bovet2005understanding}.  But in practice, for Big Data analytics
systems, the layered, modular approach has won out.  
Our goal in this paper is to re-examine this debate in the context of distributed analytics.
We seek to synthesize 30 years
of ideas in storage management and distributed analytics
into a single system---which we call Pangea\footnote{\small{Pangea is used as the storage system of PlinyCompute~\cite{zou2018plinycompute}.}}---and to examine how well that system competes with 
layered alternatives. 
As we will show, Pangea compares  favorably with 
layered systems in terms of performance.  

Pangea represents an important alternative in the distributed analytics system design space.
Developers of high-performance distributed systems need not feel compelled
to accept the costs associated with layering.
Whether the goal is developing a high-performance distributed
machine learning platform such as TensorFlow \cite{abaditensorflow}, a distributed version of a machine learning
pipeline such as
scikit-learn \cite{pedregosa2011scikit}, a Big Data computing platform such as our own
PlinyCompute \cite{zou2018plinycompute}, or high-performance distributed query or SQL engine such as
Impala \cite{kornacker2012cloudera}. If performance is key, a monolithic base 
such as Pangea should be considered as 
an alternative to building on
a layered storage system, and also as an alternative to the all-too-commonly-chosen
option of developing a distributed storage system from scratch to achieve the best performance.

Some of the key ideas embodied 
by Pangea are: 
}

\vspace{3 pt}
\noindent
1. \emph{Different types of data should be managed simultaneously by the storage system 
and different types of data should be handled
differently.} 
Thus, we redefine the locality set abstraction in classical
relational systems~\cite{chou1986evaluation} to enable the storage manager to simultaneously
manage different data durability requirements, lifetimes, and
access patterns at runtime. We describe a paging strategy that utilizes
that information to optimize page replacement decisions based
on a dynamic priority model that utilizes the locality set abstraction.
  that utilizes the locality set abstraction.

\vspace{3 pt}
\noindent
2. \emph{Data placement and replication should be integrated within the storage system.}
In distributed storage systems such as HDFS, data are replicated for fault tolerance; data are then typically
replicated \emph{again} at higher levels with different data
placement schemes for application-specific processing. \textcolor{black}{In Pangea, data may be replicated
using different partitioning strategies, and then such information is available to an external query optimizer so that a particular copy will be chosen at runtime.}  Pushing this functionality into a single monolithic system means that applications can
share and re-use such physical organizations, and it obviates the need to store even more copies to facilitate
recovery for node failures. 

\vspace{3 pt}
\noindent
3. \emph{Computational services should be pushed into the storage system.}  Thus, Pangea ships with a set of \emph{services} 
that provide an application designer
with efficient implementations of batch or operational processing operations such as sequential
  read/write, shuffle, broadcast, hash map construction, hash aggregation, and so on. 
Moving these services inside of Pangea allows job input/output
  data and ephemeral execution
  data buffered/cached all be managed within the same system. Those
  services can also bridge the knowledge gap between storage and
  applications so that storage can understand various
  application-specific performance implications for paging and data placement.

\vspace{5pt}
\noindent
\textbf{Our Contributions.}
This paper proposes a monolithic
architecture for a storage system.
Key contributions
are:

\vspace{-\topsep}
\begin{itemize}[leftmargin=*]
\setlength{\itemsep}{0pt}
\item We describe the design and implementation of Pangea, a monolithic distributed storage system, designed
to avoid the problems of heavily layered systems.
Pangea is implemented as more
  than $20,000$ lines of C++ code. 
\item Pangea presents a novel storage design that consolidates data that would typically be stored using multiple,
  redundant copies within multiple layers into 
  a locality set
  abstraction (redefined based on DBMIN~\cite{chou1986evaluation}) within a single layer. A locality set can be aware of rich
  data semantics and use such information for multiple purposes, such as replication
  and page eviction.
\item We conduct a detailed performance evaluation and comparison of
  Pangea to other related systems for applications such as $k$-means
  and TPC-H and various Pangea services. \eat{The results show that Pangea
  outperforms the
  virtual OS memory system, the OS file system, HDFS, Alluxio,
  Redis and Spark by a factor between two and 50 times. }For $k$-means, Pangea achieves more than a six times speedup compared with layering-based systems like Spark. For the TPC-H benchmark, Pangea achieves up to a twenty times speedup compared with Spark. For various Pangea services, Pangea achieves up to a fifty times speedup compared with related systems.
 
\end{itemize}
\vspace{-\topsep}

\section{Related Work}
\label{sec:survey}

{\color{black} Pangea is a monolithic system that encompasses many different functionalities: a distributed file 
system, a memory management system, and various distributed services.  Many recent papers and
projects have examined these topics in the context of Big Data and
data analytics.

Distributed file systems (DFSs), such as the Google File System~\cite{ghemawat2003google}, the
Hadoop Distributed File System~\cite{borthakur2008hdfs}, Microsoft
Cosmos~\cite{chaiken2008scope}, and IBM GPFS-SNC~\cite{gupta2011gpfs} provide scalable and
fault-tolerant persistent file storage. 
Distributed Object Storage (DOS) systems such as Amazon S3~\cite{s3}, Google Cloud Storage~\cite{GoogleCouldStorage}, Microsoft
Azure Cloud Storage~\cite{calder2011windows}, OpenStack Swift~\cite{arnold2014openstack}, Ceph~\cite{weil2006ceph} also provide storage for 
persistent objects. 
These typically provide simple operations such as 
\texttt{select} and \texttt{aggregate} which resemble Pangea's services,
but as Pangea is meant to be a general-purpose substrate for
building distributed analytics systems, its operations (shuffle, hash aggregation, join map construction,
etc.) tend to be lower-level.

Many existing systems include similar functionality to that offered by Pangea, 
though as a monolithic framework, 
Pangea includes a wider range of functionalities in a single system and subsumes these
narrower systems.
For example, in-memory file systems such as
Alluxio~\cite{li2014tachyon, li2018alluxio} can be deployed
on top of DFS and DOS to allow disk file blocks or objects to be
cached in memory and accessible to different upper-layer cluster
computing systems. Ignite~\cite{ignite} can store Spark data as SharedRDDs and cache it in Ignite'system.
The built-in memory management in frameworks such as Spark (with RDD cache~\cite{zaharia2012resilient}
and Datasets/DataFrames~\cite{tungsten,
  armbrust2015spark} and extensions such as 
Deca~\cite{lu2016lifetime}
are responsible for loading and caching input data.
Workload-aware storage systems, such as
BAD-FS~\cite{bent2004explicit}, also include a subset of Pangea's functionalities.
BAD-FS includes a job scheduler that
makes workload-specific decisions for consistency, caching and
replication. Data for each workload is managed separately and 
most of optimizations are for the cluster-to-cluster
scenario in wide-area network. 

\eat{
Tab.~\ref{tab:storage} provides a comparison of several representative storage solutions for Big Data systems.}

\vspace{3pt}
Among its other components, Pangea includes a paging system.
There has been extensive work
on page replacement algorithms such as LRU-K~\cite{o1993lru},
DBMIN~\cite{chou1986evaluation}, LRFU~\cite{lee2001lrfu}, MQ~\cite{zhou2001multi} and so
on.
As we will describe in Sec.~\ref{sec:abstraction} in detail, Pangea
borrows and extends the idea of \emph{locality sets} from DBMIN. 
Other algorithms and systems mainly consider recency, reference distance, frequency, and lifetime,
which is effective when processing single type of data as in
traditional relational database buffer pool or RDD cache, but can be
less efficient for managing multiple types of data in Big Data analytics, such as large volume of intermediate
data produced during computations (for example, during
hash aggregations and joins).
GreedyDual~\cite{cao1997cost, young1994thek} is a widely used cache
replacement framework which associates a numerical value reflecting
the desirability of keeping an object in cache. Objects are kept in
cache or replaced based on these numerical values. Independent reference model (IRM)~\cite{fagin1978efficient} and its extensions~\cite{garetto2015efficient, fonseca2002intrinsic, jaleel2010high, wu2013studying} can model cache references to different pages in order to estimate the hit/miss ratio, which are orthogonal with this work and can be applied to model and evaluate our proposed approach in the future.

There has been work on understanding access patterns of
applications and mapping them to the right storage
layer. \texttt{fadvise} and
\texttt{madvise} allow users to specify file access
  patterns. Self-learning storage systems such as
  ABLE~\cite{ellard2003attribute, mesnier2004file} predict access
  patterns based on generic file system
  traces.

Pangea also manages multiple partitionings and uses those for distributed replication and failure recovery.  Some efforts consider
data partitioning in object-oriented systems
(CoHadoop~\cite{eltabakh2011cohadoop} and
Hadoop++~\cite{dittrich2010hadoop++}) and others in SQL-based systems 
(SCOPE~\cite{zhou2012advanced}, SQLServer~\cite{agrawal2004integrating}, DB2~\cite{rao2002automating} and so on). These mainly rely on standalone failure recovery mechanisms: replicating each block to several nodes~\cite{borthakur2008hdfs}, using erasure code to store parity blocks~\cite{sathiamoorthy2013xoring}, and so on. So each partitioning may incur additional redundancy in terms of replicated blocks or parity blocks. 
C-Store~\cite{stonebraker2005c} is a column store for relational data
that maintains K-safety; multiple projections and join indexes for the
same data are maintained, so that K sites can fail and the system can still maintain
transactional consistency. }


 \section{Overview of Pangea}
\label{sec:overview}
Pangea is designed to manage all data---both intermediate 
and long-lived data, and their buffer/caching, and placement---in a monolithic storage
system. 
We begin by detailing the fundamental problem that makes devising a unified, ``Swiss army knife'' system
difficult: the disparate data types with different properties that an analytics system
will encounter.

\subsection{Disparate Data Types and Key Properties}
\label{sec:factors}
In the context of buffer pool management and file caching,
devising fair and efficient policies for allocating memory among
multiple competing datasets is always a difficult
problem~\cite{
  cao1996implementation, chou1986evaluation}. Unfortunately, the
problem is \emph{even more difficult} in the
context of analytics processing, due to the fact that there are more types of data
that the system may need to manage (consider the simple example of $k$-means clustering,
as illustrated in
Fig.~\ref{fig:kmeans}).  

In analytics processing, data can be coarse-grainedly categorized into
following: (1) {\it User data}, which is the ultimate input and output of
batch processing
applications. (2) {\it Job data}, which is short-lived intermediate
data resulting from transformation pipelines running over user data.
Then there is intermediate data that exists for a short time within
the transformations, as they are executed, which includes
(3) {\it Shuffle data} that is moved across workers (such as between map
workers and reduce workers in a MapReduce job) and (4) {\it Hash data} that
consists of a hash table and key-value pairs, as used in hash-based
aggregation and join operations. Other types of short-lived intermediate data exist as well---this is generally termed
\emph{execution
data}. Key differences in those data types (and also in different datasets of
the same type) are in following properties:

\setlength{\belowcaptionskip}{-10pt}
\begin{figure}[t]
\centering
\includegraphics[width=3.5in]{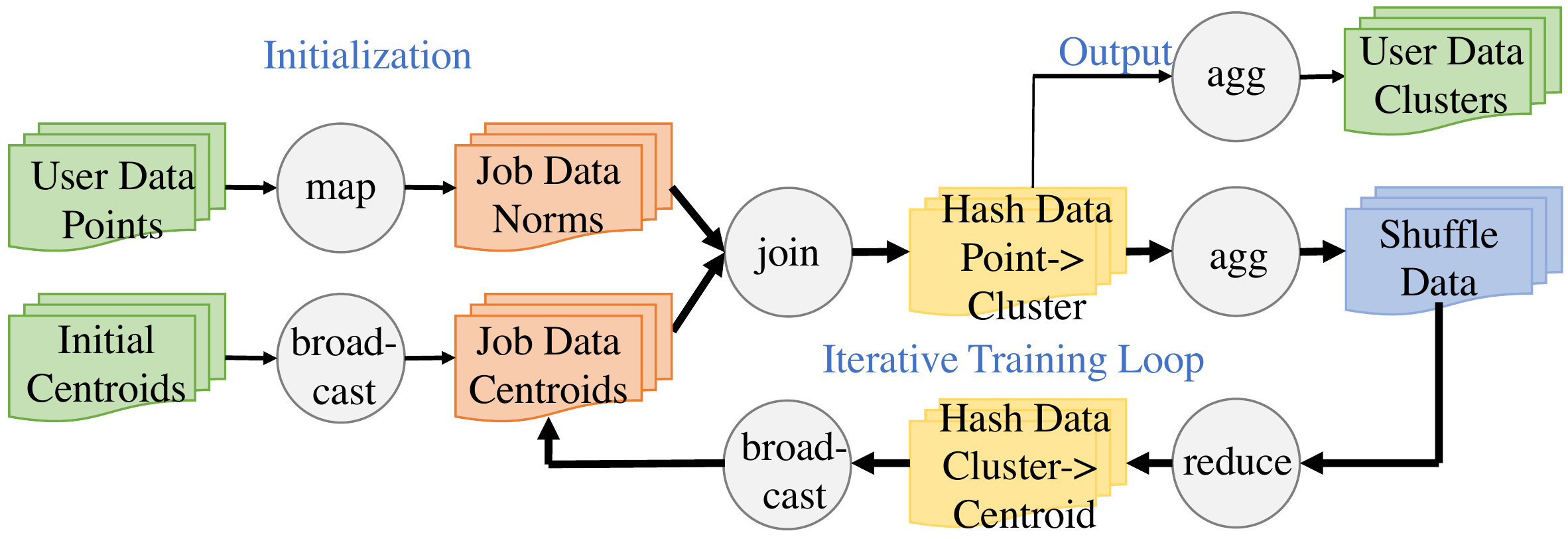}
\setlength{\belowcaptionskip}{-10pt}
\caption{ A $k$-means Example. Even a computation as simple as $k$-means
produces user data, intermediate job data, as well as execution data (the local vector of 
centroids that must be shuffled, for example).}
\label{fig:kmeans}
\end{figure}

\vspace{5 pt}
\noindent
\textbf{Durability requirements.} 
User data is read repeatedly by various applications, so it
needs to be persisted immediately once created (we call this a \texttt {write-through} requirement). However, job
data and execution data are transient, intermediate data that do not require
persistence. If transient data is being evicted from memory when its lifetime hasn't
expired, it needs to be spilled to disk (which we call \texttt {write-back}). \texttt{write-back} data
should generally have a higher priority for being kept in memory than
\texttt{write-through} data, because evicting write-back
data incurs additional I/O cost.

\vspace{3pt}
\noindent
\textbf{Data lifetimes.} In
iterative computations, the input data are needed across
multiple iterations. Shuffle data may span two job stages (e.g.
map and reduce), while aggregation data lives as long as an aggregation
is being performed.
Data that will not be accessed should be evicted
as soon as their lifetimes expire. 

\vspace{3pt}
\noindent
\textbf{Current operations.} Evicting data that has just been written
should be avoided if other things are equal, 
because such data tends to be read soon. For example, in a query execution flow that consists of multiple job stages, the input and output of a job stage may have similar durability requirements, lifetime, access recency and so on, but the output is more likely to be accessed by next job stages than the input.

\vspace{3pt}
\noindent
\textbf{Access patterns.} 
Different operators and their associated data types
often exhibit disparate access patterns. For example, a map operator processes each data
element in the input dataset, so eviction
of data elements that haven't yet
been processed should be avoided. In contrast, an aggregation
operator needs to insert each data element of input data into a hash
table, which is often randomly accessed, and should have high priority in staying in memory.

\vspace{3pt}
\noindent
\textbf{Temporal recency.} 
Datasets that have been accessed more recently are more likely
to be reused, e.g. recurring services~\cite{jyothi2016morpheus,jindal2018computation,chen2012interactive}.

\subsection {The Locality Set Abstraction}
\label{sec:abstraction}

In a layered system, data within a layer (e.g. a user file system, or
a memory pool in computing platform)
tend to be relatively homogeneous. For example, HDFS and Alluxio only store
user data for various applications; the Spark RDD cache stores user data and job data only for the current
application that is in processing; the Spark execution memory pool only
stores execution data for the current application. Pangea's task is more difficult, because it manages all 
data within a single layer---data exhibiting different access patterns, data
sources, and lifetimes are present. 

To facilitate management of
disparate data types, we 
borrow and update the \emph{locality set} abstraction, which was first proposed in
DBMIN~\cite{chou1986evaluation} as the query locality set model (QLSM) for buffer
management in classical relational systems. QLSM mainly defines a set of
access patterns; and for each access pattern, there is a predefined
eviction policy such as MRU, LRU, and so on, and also a predefined
algorithm to derive the desired size of locality set, where a locality
set is defined
as the set of buffer pool pages for a file instance. DBMIN's buffer
pool management strategy is based on the desired size and eviction policy of each
locality set. When the size of a locality set is larger than its
desired size, a page will be chosen and evicted based on the eviction
policy associated with this locality set.

{\color{black}
While DBMIN's locality set idea serves as a building block for Pangea, DBMIN was
first proposed more than 30 years ago for transaction processing, at a time
when modern analytics workloads did not exist.  As such, some of DBMIN's basic assumptions need to be updated.
Most problematic
is the implicit assumption in DBMIN that all locality sets correspond to persistent database tables.
As such, DBMIN does not 
consider durability requirements and assumes all data should be persisted to disk---a fine
assumption in transaction processing, but not 
reasonable for modern analytics workloads that frequently produce large volumes of 
transient intermediate data.  
DBMIN requires a maximum number of buffer pages (i.e. desired size) on a per file-and-query basis to be known (or guessed) 
beforehand, which is again not
reasonable for intermediate datasets produced as the result of running opaque user-defined functions on data, as is
standard in modern data analytics.  
If the total number of estimated buffers exceeds available memory, new requests are blocked, which is again unreasonable
on modern analytics workloads.}

\setlength{\belowcaptionskip}{-10pt}
\begin{table}[t]
\centering
\scriptsize
\caption{\label{tab:attributes} Some Locality Set Attributes.}
\begin{tabular}{|p{2.8cm}|p{5cm}|} \hline
Attribute&Supported values\\\hline \hline
\texttt{DurabilityType}&\texttt{write-back}, \texttt{write-through}\\ \hline
\texttt{WritingPattern}&\texttt{sequential-write}, \texttt{concurrent-write}, \texttt{random-mutable-write}\\ \hline
\texttt{Location}&\texttt{pinned}, \texttt{unpinned} \\
\hline
\texttt{ReadingPattern}&\texttt{sequential-read}, \texttt{random-read} \\
\hline
\texttt{Lifetime}& \texttt{lifetime-ended}, \texttt{alive}\\ \hline
\texttt{CurrentOperation}&\texttt{read}, \texttt{write}, \texttt{read-and-write}, \texttt{none} \\ \hline
\texttt{AccessRecency}& \texttt{sequence id for last access}\\ \hline
\end{tabular}
\end{table}

{\color{black} Thus, we update some of the ideas in DBMIN.}
A Pangea locality set
is simply a set of 
pages (or blocks) associated with one dataset that are used by an application in a uniform way and are
distributed across a cluster of nodes. 

{\color{black} In Pangea, there is no hard partitioning of the buffer pool to 
the different locality sets (which, for transient data created via UDFs, would require solving difficult
estimation problems, such as guessing the size of the locality set).  
All Pangea locality sets share the same buffer pool (see Sec.~\ref{sec:bufferpool}) and unified
eviction policy (see Sec.~\ref{sec:paging}).}

In Pangea, all pages in one locality set must have the same size, which can be configured when creating the locality set.
Data organization in a page is flexible, and each page can represent a chunk of relational
data, or a container (e.g. vector, hash-map and so on) of 
objects. Pages from a locality set may be stored on disk.  However, unlike DBMIN, it is not necessary for each
page of a locality set to have an image in an associated file. In
Pangea, a locality set page can
reside in disk, in memory, or both, because 
transient data (job data and execution data) are also stored in locality sets; such 
locality sets may have had only a fraction (or none) of their pages
on disk.

Depending on the application requirements, different locality sets must be managed differently.
To achieve this, different with DBMIN, Pangea locality sets are given
a set of \emph{tags}, or attributes, either by an application 
or automatically by inference.  The various attribute categories and
their supported values are listed in
Table~\ref{tab:attributes}.
The attributes generally correspond 
to the key factors as identified in Sec.~\ref{sec:factors} and are
first considered as locality set attributes.

In analytics processing, there are typically
three writing patterns: \texttt{sequential-write} where immutable
(write-once) data is written to a page sequentially; \texttt {concurrent-write}
where multiple concurrent data streams are written to one page
(write-once), e.g. to support creation of shuffle data; \texttt {random-mutable-write} where data can be dynamically allocated,
modified, and deallocated in a page (write-many), e.g. to support aggregation,
pipeline processing, or join. There are two reading
patterns: \texttt {sequential-read} and
shuffle data; and \texttt{random-read} for data such as hash data. 

\vspace{3 pt}
\noindent
\textbf{Determining attributes.}
Pangea provides various \emph{services} to
read and write locality sets. In Pangea, attributes such as \texttt{ReadingPattern}, \texttt{WritingPattern}
and \texttt{CurrentOperation} are automatically
determined at runtime through invocations of services, since
each service exhibits specific writing and 
reading patterns. For example, the sequential read and write services
exhibit the \texttt{sequential-read} and the \texttt{sequential-write} pattern, respectively. The
shuffle service exhibits the \texttt{concurrent-write} pattern,
and the hash service exhibits both the \texttt{random-mutable-write}
and \texttt{random-read} patterns.  Thus, when (for example)
an application associates the locality set with a
sequential allocator that provides the sequential write service, then \texttt{WritingPattern} must be
\texttt{sequential-write} and \texttt{CurrentOperation} must be \texttt{write}; if an
application associates the locality set with a sequential
iterator that provides the sequential read service, then \texttt{ReadingPattern} must be \texttt{sequential-read}, 
and  \texttt{CurrentOperation} must be \texttt{read}.

\vspace{3pt}
Here are some examples of using Pangea locality sets.

\begin{codesmall}
//create a set
LocalitySet myData = createSet("data");
//add single object (sequential write)
myData.addObject(myObject);
//add a vector of objects (sequential write)
myData.addData(myVec);
//sequential read
vector<PageIteratorPtr> * iters = 
  myData.getPageIterators(numThreads);
for (int i = 0; i < iters->size(); i++) {
  // to start worker threads to read pages
  runWork(iters->at(i), myWorkFunc); 
}
//create a set for storing shuffled data
LocalitySet shuffledData = createSet("shuffled");
//invoke shuffle service in one worker 
while((PagePtr page = iter->next())) {
 ObjectIteratorPtr objIter 
   = getObjectIterator(page);
 while((RecordPtr record = objIter->next())) {
   PartitionID partitionId = 
     hash(udfGetKey(record));
   VirtualShuffleBufferPtr buffer = shuffledData.
     getVirtualShuffleBuffer(workerId, partitionId);
    buffer->addObject(record); 
  }
}
\end{codesmall}

More code examples for using locality set and invocation of services are
presented in Sec.~\ref{sec:partition} and Sec.~\ref{sec:services}.

\vspace{3pt}
\noindent
\textbf{Heterogeneous replication.} 
In a monolithic system like Pangea, a locality set can have multiple replicas to do double-duty
and facilitate \emph{both} recovery and computational savings due to the ability to provide
multiple physical data organizations.  Furthermore, replicas in Pangea are visible and usable
by all applications running on top of Pangea.  
Applications or a data placement optimizer can apply different partition schemes and page
sizes to a locality set to generate new locality sets, and register those replicas with the same replica group. This replica and partitioning information (along with
other useful attributes of the data)
is stored in Pangea's statistics database and made available to all applications with appropriate
permissions that are running on top of Pangea for choosing replica for computation. 

\eat{
By pushing what has
traditionally been application-level data
into the storage system
(specifically, semantic information such as the partition scheme) replicas can
be used to avoid data loss at the same time they are used to improve
application performance. For example, to execute
a query, a database running on top of Pangea does not need to load data from external
storage and then repartition or
co-partition data to optimize placement. It can select a
Pangea replica that is the best for the query execution and directly run
computations on the data, while not increasing storage size by using
the same replication factor value.} Details about data recovery using
heterogeneous replicas are described in Sec.~\ref{sec:partition}.

%

\eat{
\textbf{Jia, I changed your discussion of pinning.  Pinning has nothing to do with data lifetime.  A pinned page is one that we don't want
the system to remove from the buffer; an unpinned one is a page that can be removed, but it says nothing about the page's lifetime having ended.
The page could still be useful, it just need not be buffered in RAM.}}

\subsection{System Architecture}



The Pangea system has 
five components, which we describe at a high level now and will be described in more
detail subsequently.

\vspace{5 pt}
\noindent
\textbf{
(1) The distributed file system.}  In Pangea file system, a file
instance is associated with one locality set and consists of a
sequence of pages that are on-disk images of all, a portion or even
no pages in the
associated locality set. Those pages can be distributed
across disks in multiple nodes. 
\eat {
When flushing pages from their in-memory images to disks,  all pages of one ``locality set'' will be written to the same file instance.}

\vspace{3 pt}
\noindent
\textbf{
(2) The buffer pool.} Unlike other distributed storage systems, on each
node Pangea utilizes a unified buffer pool to manage both user data
and execution data.
The buffer pool manages most or
all of the RAM that is available to Pangea and the applications
that are running on top of Pangea, in the form of a large region of shared memory. 
The idea is that applications rely
on Pangea to collectively manage RAM for them.

\vspace{3 pt}
\noindent
\textbf{
(3) The paging system.}
The paging system is responsible for evicting pages from
the buffer pool to make room for allocating new pages. 
It maintains a dynamic priority model that orders
locality sets according to their durability requirements, lifetimes,
access patterns, access recency and so on. For each
locality set, a paging strategy is automatically selected based on its
access patterns. When Pangea needs more RAM, it
finds the locality set with the lowest priority and uses its selected strategy
to evict one or more victim pages from the locality set. 
 Details are described in Sec.~\ref{sec:paging}.

\vspace{3 pt}
\noindent
\textbf{
(4) Distributed data placement system.}
Each locality set can have multiple replicas which allow for both recovery and computational 
efficiency. Each replica may have different physical properties,
so that the replica with the best physical organization can be selected for a given computation. 
The data placement system manages
partition, replication and recovery of
locality sets and is described in Sec.~\ref{sec:partition}.

\vspace{5 pt}
\noindent
\textbf{
(5) Distributed services.} To realize the benefits of Pangea's unified storage architecture,
applications need to entrust all of their datasets (including user data, job
data, shuffle data, aggregation data and so on) to Pangea. To facilitate this,
Pangea provides \emph{services} to the applications that run on top of Pangea. The \emph{sequential read/write service}
allows each of multiple threads to use a
sequential allocator to write to a separate page in a locality set. Each worker can then use a provided concurrent page iterator to
scan a subset of pages in the locality set. The \emph{shuffle service} allows multiple writers to
write to the same page using a concurrent
allocator, so that multiple data streams for the same shuffle
partition can be combined into a single locality set. The
\emph{hash service} allows a locality set to be allocated as a
key-value hash table,
through a dynamic 
allocator. Details are discussed in
Sec.~\ref{sec:services}.

\eat{
Different services write to a 
page that is allocated from the buffer pool using different writing patterns. Pangea provides
a unique secondary allocator for each writing pattern that makes it possible for the
application to perform sub-allocations on the page, depending upon application needs.
For example, a locality set marked as \texttt{random-mutable-write} can access
a dynamic memory allocator that transforms a page into free-lists of sub-pages and
allocates buffers from the free-lists. A locality set marked as \texttt{sequential-write}
is given a simple but fast sequential allocator that writes immutable data sequentially to a page's memory.
A locality set marked as \texttt{concurrent-write} 
is provided with a
concurrent allocator that allows multiple threads to write to a page
in parallel.

Various services are then built upon this 
two-level allocation framework.} 

\vspace{3 pt}
\noindent
\textbf{
Deployment and Security Considerations.} The Pangea distributed system
consists of one light-weight manager node responsible for accepting
user applications, maintaining statistics and etc., and many worker
nodes that run the functionalities of above five components. 

For cloud deployment, Pangea ensures
security by delegating authority to remote processes
through the use of public-key. We require the users must
be assigned a valid public-private key pair in the deploying
cluster. Then the user must submit the private key when
bootstrapping the system. Under the hood, in the initialization stage,
the Pangea manager node relies on
this user submitted private key to access workers for collecting system
information. A non-valid key will cause the whole system \textcolor{black}{to terminate}.

\section {The Distributed File System}
\label {sec:filesystem}

In keeping with our ``no layering'' mantra, 
to avoid redundant data copying between a cache layer (e.g. Spark RDD,
Alluxio) and a file system layer (e.g. the HDFS, or the OS file
system), on each worker node, a Pangea process
contains 
a user-level file system which uses the Pangea
buffer pool (described in Sec.~\ref{sec:bufferpool}) to buffer reads and writes. All reads/writes are implemented via direct I/O to
bypass the OS buffer cache.

In Pangea, a distributed file instance that is associated with one locality set is implemented using one Pangea
data file and
one Pangea meta file on each worker node.
On each worker node, a Pangea data file instance is chunked into fixed size pages.
Depending on user-selected settings, a Pangea data file instance can be automatically distributed across multiple
disk drives on its worker node.  The set of pages allocated to each disk drive can be mapped to a physical disk file. The Pangea meta
file is simply a physical disk file used to index each page's location and
offset. 

A centralized Pangea manager shared by the distributed file system and
other distributed components manages locality set metadata (such as database name, dataset name, page size,
data attributes, partition scheme, replica group and so on). Metadata for each page are stored in the
Pangea meta files at each worker node. Compared to HDFS (where locations are stored for each block at the name node) the Pangea manager stores considerably less meta data.

When reading a page, Pangea first checks the buffer pool to see whether
the page is already cached in memory. If the page is not present in
the buffer pool, the page
needs to be cached first. 
When writing, depending upon a locality set's \texttt{DurabilityType}, there are
two durability settings available: (1) \texttt{write-through}, where
each page, once written, will be cached and also persisted to disk; and (2) \texttt{write-back},
so that a dirty page is first cached in the buffer pool, and
written to disk only if there is no free space in the buffer pool and the page has been chosen
to be evicted. 
User data is often
configured as write-through, while transient job data and execution data is often configured
as write-back. 

\section{The Buffer Pool}
\label{sec:bufferpool}

\begin{figure}
\centering
\includegraphics[width=3.5in]{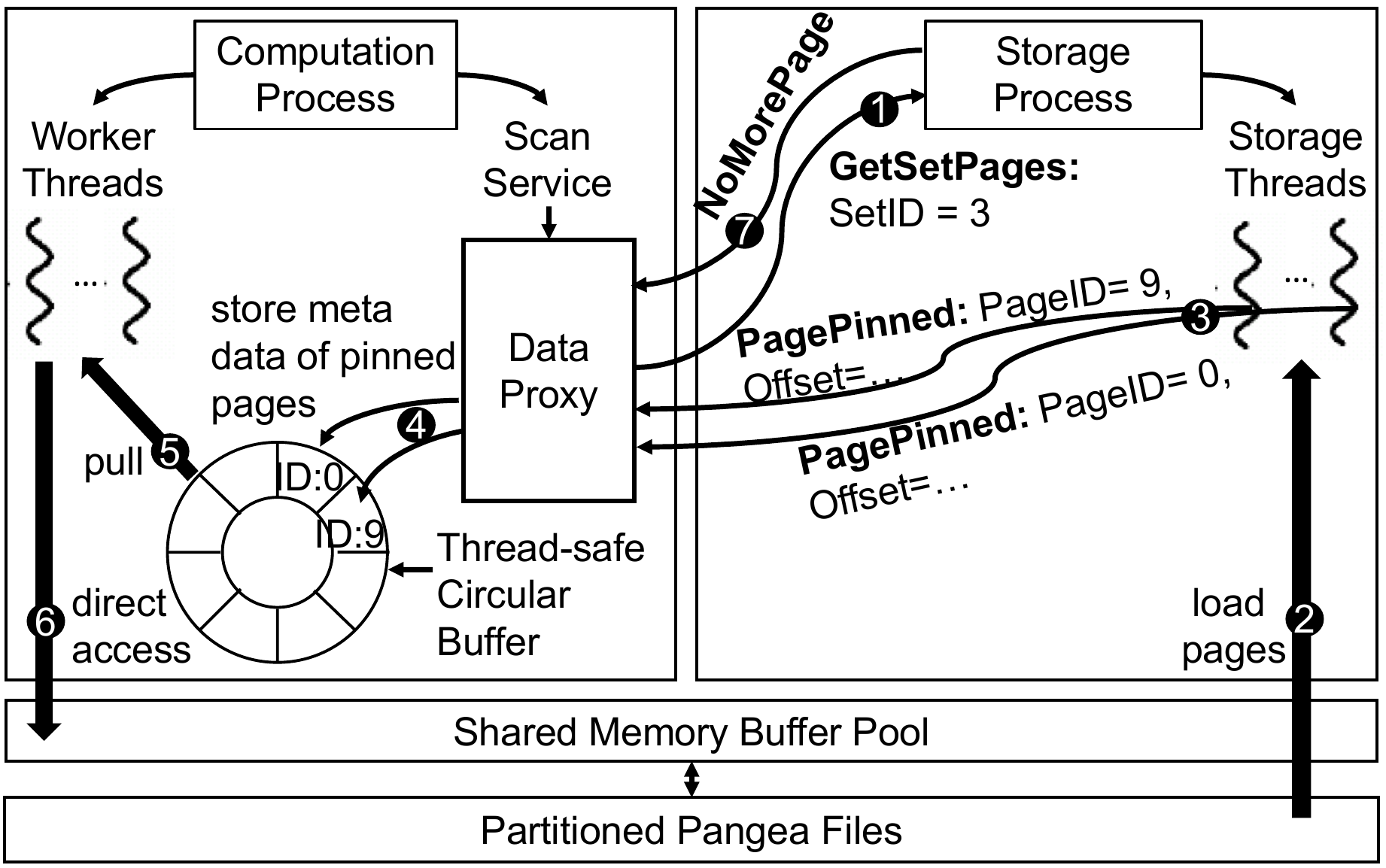}
\caption{\textcolor{black}{Data scan: long living worker threads pull page meta data from a circular buffer and access page data through shared memory.}}
\label{fig:scan}
\end{figure}

Separate buffering mechanisms partitioned across the various software layers found in today's 
analytics installations make it difficult for the various buffering software to coordinate and 
assign resources in an optimal way.\eat{For example, Spark, when processing a large volume of execution data
(cached in its execution pool),
is configured to evict only evict 50\% of cached RDD data (cached
in storage pool), which is
controlled by a pre-defined storageFraction parameter.}\eat{ \textbf{Jia, I removed your Spark example here, because it does not illustrate the point.
Spark is a single software, so they could fix the problem you mention... this is an implementation issue in Spark, not an architectural one.}} 
To sidestep this, Pangea
caches all data 
in one
unified buffer pool that is used by all applications utilizing Pangea.

On each node, Pangea uses anonymous \texttt{mmap} system call to allocate a large chunk of
shared memory as a buffer pool in a storage process.  Computation
process that has multiple worker threads
accessing the data concurrently is forked by Pangea and given access
to the shared memory.

Those computation threads coordinate
read/write access to the buffer pool with Pangea via a data proxy based on sockets, but the actual data served are communicated via
shared memory. 
In this way, data written to a buffer pool page are visible to computations
immediately, without any copy and moving overhead. A significant
fraction of the typical serialization/deserialization overhead present
in current analytics systems can also
be avoided for accessing local, in-memory data. 

All information required to access and manage a buffered page, such as
the the page offset in shared memory, 
is communicated through the network. We implement various message-based
communication protocols directly on top of TCP/IP for the computation
framework to exchange page location
information with Pangea's storage facilities.

Taking sequential read\textcolor{black}{, as shown in Fig.~\ref{fig:scan},} for example, the data proxy in the computation
process first sends a \texttt{GetSetPages} message to
the storage process. Then the storage process starts multiple threads \texttt{pin}ning the pages in the locality set
to be scanned, and the meta data (e.g. relative offset in the shared
memory) for each page that is \texttt{pin}ned to the buffer pool is sent to the
computation process via socket. The data proxy then put those meta
data into a circular buffer that is thread-safe, from which multiple
computation threads can fetch the meta data for a page at a time, and
access that page through shared memory for running computations. 

\vspace{3pt}
Therefore, in the computation process, the threading
model supported by Pangea,
is quite different with the ``waves of tasks'' model used in Spark
and Hadoop where a task thread will be scheduled for each block of
data, and many such small tasks will execute concurrently as a
``wave''. 

Instead, in applications and computation
frameworks~\cite{zou2018plinycompute} built on Pangea, for executing a
job stage, the computation process start
multiple worker threads, which are long living and do not terminate until all input data
for the job stage has been executed. Then in a loop, each worker
thread pulls page meta data from the concurrent circular buffer as
described above.

Therefore, in Pangea, there is no need to consider the ``all-or-nothing''
property of ``wave''-based concurrent task execution, trying to have all or none of the inputs for tasks
in the same wave cached, as proposed in memory caching system
for data-intensive parallel jobs such as PACMan~\cite{ananthanarayanan2012pacman}.

\vspace{3pt}
To write output to a page, the data proxy sends a \texttt{PinPage}
message to the storage process, which then pins a page
in the buffer pool, and send back the meta data of the page to the
computation.

The buffer pool on each node uses a dynamic pool-based memory allocator to allocate
pages with various sizes from the same
shared memory pool. \textcolor{black}{Pangea supports two main pool-based memory allocators: the Memcached slab allocator~\cite{fitzpatrick2004distributed} and the
two-level segregated fit (TLSF) memory
allocator~\cite{masmano2004tlsf}. We use TLSF by default
because it is more space-efficient for allocating variable-sized pages from the shared memory.}

A hash table is used to map the page identifier to
the page's location in the buffer pool.
Each page cached in the buffer pool can be mapped to a
page image in one locality set. Once the page must be spilled, 
the page will be appended to the
locality set's associated file instance. 

As mentioned, each page has a \texttt{pinned}/\texttt{unpinned} flag.
In addition, each page has a \texttt{dirty}/\texttt{clean} flag
to indicate whether the page has been modified. Page reference
counting is used to support
concurrent access. A service can \texttt{pin} an existing page
in a locality set for reading and writing. Once the page
is cached in the buffer pool, its flag is set to \texttt{pinned}, with its
reference count incremented. 

Once the processing on a page has finished, the service \texttt{unpin}s
the page and the reference count is decremented. When the buffer pool has no free pages to allocate for an incoming
pinning request, the paging system will evict one or more
unpinned pages and recycle their memory. Before evicting
an unpinned page that is marked as ``dirty'' but is still within its
locality set's lifetime, we need to make sure
that all the
changes are written back to the Pangea file system first. We
describe more detail about paging in Sec.~\ref{sec:paging}.

\section{Paging System}
\label{sec:paging}

{\color{black}
Given a page pinning request when there are no free pages in
the buffer pool, the paging system will select a locality set whose next page-to-be-evicted
has the
lowest priority and ask that locality set to evict one or more pages
based on this locality
set's selected paging strategy. Each locality set has its
own paging policy, chosen to match its attributes.  Pangea selects MRU as
the paging strategy for locality sets labeled
\texttt{sequential-write}, \texttt{concurrent-write} and \texttt{sequential-read}, and LRU as the paging strategy
for locality sets labeled \texttt{random-mutable-write} and \texttt{random-read}. 

The number of pages evicted from the selected locality set is based on the 
\texttt{CurrentOperation} locality set
attribute. If the selected locality set is labeled
\texttt{write} or \texttt{read-and-write}, then a single victim page is selected, as evicting data that has just been written should be avoided, as explained in Sec~\ref{sec:factors}. For read-only locality sets,
10\% of the locality set is evicted, as we have found that in Big Data analytics, read-only operations
tend to be well-behaved, with a lot of temporal locality.  Once an application has not read one page in a while,
it is unlikely to use any pages from the set for awhile, and a larger eviction is warranted.

The key question is how the priority of each locality set's next page-to-be-evicted's priority is computed.
Assuming for a moment that each locality set has only pages that are alive (\texttt{lifetime-ended} is false),
we choose a locality set to be the victim locality set if the next page-to-be-evicted from the locality set
has the lowest expected cost compared to the priority of the next page-to-be-evicted from all other locality sets, 
within a given time horizon $t$.  

There are two parts to this expected cost: the cost $c_w$ is the cost to write out the page, 
and the cost $c_r$ to read it again, if necessary.  Then, the overall expected cost 
of evicting a page is:
$$c_w + p_{reuse} \times c_r$$
\noindent
where $p_{reuse}$ is the probability of accessing the page within the next $t$ time ticks.
That is, no matter what, we pay a cost for evicting the page, and we may pay a cost associated
with reading the page if it is re-read at a later time.

$c_r$ can be estimated as $c_r={v_r}\times{w_r}$ where $v_r$ is the profiled time to read the page from disk; 
$w_r$ represents the penalty associated with the reading 
pattern of a locality set. Reading spilled data for the sequential read 
pattern only requires reading the page to memory, so $w_r=1$. But reading spilled
data that has a random reading pattern (\texttt{random-read}) incurs a higher cost ($w_r>1$),
because reading such data requires reconstruction of the hash map and re-aggregation of the spilled data. 

$c_w$ is determined by the locality set's durability requirement and can be computed as $c_w=\frac{d}{v_w}$. $v_w$ the time to write the page to disk (also collected via profiling) and $d=1$ for the \texttt{write-back} requirement, and $d = 0$ for \texttt{write-through}.

$p_{reuse}$, which is the probability that the page is reused in the next $t$ time ticks,
is a bit more complicated to estimate.  $p_{reuse}$ for a page is computed from the page's $\lambda$ value, where $\lambda$
is the rate (per time tick) at which the page is referenced.
If we model the arrival time of the next reference to each page as a Poisson point process \cite{kleinrock1976queueing}, then the probability that the page is referenced in the next $t$ time 
ticks is $1 - e^{-\lambda t}$ (this follows from the cumulative density of the exponential distribution, which models the time-until-next arrival for a Poisson point process).

There are a number of ways that $\lambda$ for a page can be estimated.
We can collect the number of references $n$ to the page in the 
last $t'$ time ticks, and estimate the rate of references per time tick as $\lambda \approx n/t'$.
This quantity is a bit difficult to deal with in practice, however.  It requires storing multiple 
references to each page, maintained over a sliding time window.  

$\lambda$ can also be estimated from the time since the last reference, which is what we use in our Pangea implementation.  
If a page was last referenced at time tick $t_{ref}$ and the current time tick is $t_{now}$, the number of references
since the beginning of time can be estimated as $t_{now} / (t_{now} - t_{ref})$; dividing by $t_{now}$ to get the number
of references per time tick gives $\lambda \approx 1 / (t_{now} - t_{ref})$; that is, $\lambda$ is the inverse of the time-to-last
reference of the page.\footnote{\small {The inverse of the page's reference distance can also be seen as yet another reasonable
estimate for $\lambda$, as this
effectively replaces $t_{now} - t_{ref}$ with the page's
last observed between-reference time as an estimate for the expected time interval between page references.  We choose the time
since last reference, however, as it requires only a single reference to be valid.}}

Finally, note that the previous discussion assumes that 
\texttt{lifetime-ended} is false for each locality set.  If there are one or more locality sets where
\texttt{lifetime-ended} is true, these are always chosen for eviction first, again according to the minimum expected
cost of evicting a page from the locality set.

\vspace{5 pt}
\noindent \textbf{A note on rate vs. probability.}
There is a strong relationship between using $p_{reuse}$ computed via an exponential distribution
with a time horizon of $t = 1$, 
and simply weighting 
a page's read cost $c_r$ by $\lambda$ (the inverse of the time since last reference in the case of Pangea).
In fact, the latter is a linear approximation to the former.
If one approximates the exponential computation of $p_{reuse}$ with a linear function 
(a first-degree Taylor series approximation of the exponential function about the point $\lambda' = 0$), we have:
\begin{align}
p_{reuse} &= 1 - e^{-\lambda t} 
          \approx 1 - e^{-\lambda' t} + te^{-\lambda' t}(\lambda - \lambda') \nonumber \\
          &= 1 - e^{0} + te^{0}\lambda 
          = t\lambda = \lambda \nonumber
\end{align}
}

\section{Data Placement}
\label{sec:partition}



Pangea's monolithic architecture allows it to use replication to perform double-duty: to provide for fault tolerance (as in HDFS), \emph{and} to provide for computational efficiency by allowing for multiple physical data organizations.

Appropriate data partitioning (such as co-partitioning of related data on the same join key) can avoid shuffling across the network, and speed up operations such as joins by many times~\cite{eltabakh2011cohadoop}. 

While systems such as Spark provide similar functionality, a partitioned RDD in Spark is specific to computation and will be discarded once an application runs to completion; it can not be reused for
future runs of applications. Although a
Spark developer could materialize the repartitioned dataset as a
different HDFS or Alluxio file, there are two shortcomings: (1) HDFS
can not recognize and utilize those files for failure recovery; (2)
the process of loading data into RDD cache is
controlled by Spark task scheduler, which is optimized for locality
but doesn't guarantee locality,
thus a repartition stage at runtime is still needed before performing
a local join
that can be pipelined with other computations. As a result, a Spark application often
invokes \texttt{repartition()} to tune the split size and then applies
\texttt{partitionBy()} to tune the partitioning scheme every time the
application is executed.

In Pangea, such physical data organizations are persistent and can be
shared across applications. For example, a {\it source set}
\texttt{lineitem}  can be
partitioned into a {\it target set} \texttt{lineitem\_pt}:

\begin{codesmall}
registerClass("LineItem.so");
LocalitySet myLineItems = getSet("lineitem");
LocalitySet myReplica = 
  createSet<LineItem>("lineitem_pt");
PartitionComp<String, Lineitem> partitionComp = 
  PartitionComp(getKeyUdf);
partitionSet (myLineitems, myReplica, partitionComp);
\end{codesmall}

{\color{black} This code creates a partition computation that extracts a \texttt{String}-valued key from each \texttt{Lineitem} and uses that key to partition the contents of \texttt{myLineItems}.} Then, a user can register \texttt{lineitem\_pt} as a replica of
lineitem using a \texttt{registerReplica()} API:

\begin{codesmall}
registerReplica( myLineItems, myReplica,
 numPartitions, numNodes, partitionComp);
\end{codesmall}

{\color{black} Now, any application that finds the partitioning useful can use this new replica to perform computations. 

Under the hood, the source set and the target set are placed in the same \emph{replication 
group}. By definition, each set in a replication group contains exactly the same set of objects organized
using a different physical organization, and
an application running on top of Pangea can choose any appropriate set in the replication group, based
on the desired physical properties of the set.
Registering multiple sets in a
replication group in this way has the added benefit of obviating the need to store
multiple copies of each object on different machines in order to allow for failure recovery.

However, having the sets in a replication group do ``double duty'' in this way requires some care.
Because the various physical organizations are chosen for computational reasons (pre-partitioning
based upon a join key to make subsequent joins faster, for example) all copies of an object
may just happen to be stored on the same machine.  We call such objects ``colliding'' objects.
Colliding objects are a problem because if the machine holding the colliding objects fails, the objects are lost.
}

In practice, however, the number of colliding objects is small.
If a transformation to a target set is random (hashing, for example), 
then the expected number of
colliding objects can be estimated as $n/k$, for $n$ objects and $k$
worker nodes. 
In our experiments using real partitioning schemes, the number of colliding objects is small. When
partitioning the TPC-H lineitem table that has $5.98$ billion \texttt{Lineitem} objects (about
$79$GB in size) using two partitioning schemes onto ten Pangea worker nodes
(on \texttt{l\_orderkey} and \texttt{l\_partkey} respectively),
there are $53.39$ million colliding objects in total. When we
partition the same \texttt{Lineitem} table to $20$ nodes,
there are only $15$ million colliding objects. When we further use $30$
nodes for the same partitioning, we find no colliding objects.

Given that the number
of colliding objects is relatively small, 
to achieve a complete recovery of lost data, we identify
and record all colliding objects at partitioning time. Then those colliding 
objects will be stored in a separate locality set, and replicated using an approach similar to 
HDFS replication. 

\textcolor{black}{
The recovery process first requires calculating the key range for all lost partitions from the failed node. Then, to recover a particular replica (referred to as the \emph{target replica}), the system arbitrarily selects another replica from the replication group (the \emph{source replica}). The system runs the target replica's partitioner on the source replica to extract the key for each object in the source replica. If a key falls in the range of lost partitions, the key and associated object are buffered and later dispatched to the location where the associated key range in the target replica is being recovered. At the same time that the source replica is being processed, all colliding objects from the replication group whose instance in the target replica have been placed on the failed node are recovered. These objects are recovered by processing the special locality set used to store the colliding objects from the replication group.} 

The  strategy can be extended to handle concurrent $r$-node
failures by separately replicating any object of which the replicas are located on fewer than $r+1$ nodes. This requires significantly more disk space (i.e. if partitioning is random, the expected ratio of such objects in $k$-node cluster is $(1-\dfrac{k\times (k-1)\times ... \times (k-r)}{k^{r+1}})$). Fortunately, since
modern analytics frameworks are usually deployed on smaller clusters
ranging in size from a few to a few dozen nodes, concurrent
multiple-node failures are the exception ~\cite{clusterSize, crotty2015architecture}.

\section{Services}
\label{sec:services}
\eat{The advantage of Pangea's non-layered approach can be fully realized if
applications use Pangea to manage all of their data: user data,
job data, shuffle data, aggregation data and so
on. To enable this, we build a service framework into Pangea.
A \emph{service} is an implementation of a data-processing
pattern within Pangea.} Pangea provides a set of services to enable various types of locality sets to be cached in one buffer on each worker node and their attributes to be learned at runtime. We now describe a few of the services
offered by Pangea.

\vspace{3pt}
\noindent
\textbf{The Sequential Read/Write Service.}
This service allows one or more threads on each worker node to read or
write data to or from a locality set. To write to a locality set
sequentially, a worker first needs to configure the locality set to
use a sequential allocator to allocate bytes from the
page's host memory sequentially for writing byte-oriented data. If a page is fully written, the storage will unpin
  the page and pin a new page in the locality set. 

 To scan a locality set using one or multiple threads on a worker node, the application first needs to obtain a set of
 concurrent page iterators from the locality set and dispatch each
 iterator to a thread, using code like the following:

\begin{codesmall}
LocalitySet myInput = getSet(setId);
//if "write-back" is not specified here,
// "write-through" is used by default.
LocalitySet myOutput = 
  createSet(setName, "write-back");
//if "sequential" is not specified here, the dynamic 
//secondary allocator will be used by default.
myOutput.setAllocationPolicy("sequential");
vector<PageIteratorPtr> * iters = 
  myInput.getPageIterators(numThreads);
for (int i = 0; i < iters->size(); i++) {
  // to start worker threads
  runWork(iters->at(i), myOutput, userfunc); 
}
\end{codesmall}

Then in each thread, we sequentially write to pages pinned in \texttt{myOutput}:

\begin{codesmall}
while((PagePtr page = iter->getNext())) {
  ObjectIteratorPtr objectIter 
      = createObjectIterator(page);
  while((RecordPtr record = objIter->next())) {
    myOutput.addObject(userfunc(record));
  }
}
\end{codesmall}
  \eat{each disk drive
  will have one producer thread to iteratively load pages from the disk's data partition
  to the buffer pool, and send the metadata
  information of loaded pages to the computation process (as
  illustrated in Fig.~\ref{fig:communication}). The computation
  process will add those metadata information to a concurrent
  queue. Then, one or multiple reading threads can fetch metadata
  information from the concurrent queue, and sequentially read the data in pages through
  shared memory.

  Synchronization overhead of concurrent queue is low in above
  process, mainly because we use comparatively large page size.}

\vspace{3pt}
\noindent
\textbf{Virtual Shuffle Buffer/Shuffle Service.}
For shuffling, all data elements dispatched to the same partition
need to be grouped in the same locality set, and we create one locality set
for each partition. 

It is important to allow multiple shuffle writing threads to write data elements belonging to
the same partition to
one page concurrently, to reduce batch latency and memory
footprint. Thus, we use a secondary, small
page allocator that first pins a page in the partition's locality set, then 
dynamically splits small pages (of several megabytes) from a large
page, and allocates small pages to multiple threads. Once all
small pages are fully written, the small page allocator unpins this page and allocates a
new page from the buffer pool for splitting and allocating small pages. 

To allow threads to access small pages transparently, we offer a
virtual shuffle buffer abstraction. Each shuffle writer allocates one virtual shuffle buffer for each partition. A virtual shuffle buffer contains
a pointer to the small page allocator that is responsible for the partition's
locality set,
and also the offset in the small page that is currently
in use by its thread. 
Then, each partition's locality set
  can be read via the
 sequential read service.
\eat{
\begin{figure}
\centering
\includegraphics[width=3.2in]{fig/virtual-shuffle-buffer}
\caption{Shuffle Data Management in Buffer Pool.}
\label{fig:virtual-shuffle-buffer}
\end{figure}}

\eat{
  Each shuffle
  writer (e.g. Map task) need first register with this service, and
  get handles to
  a set of virtual shuffle buffers, with each containing a handle to the locality set that is
  corresponding to one shuffle reader (e.g. a Reducer). Then the
  writer can write immutable bytes to those shuffle
  buffers. As described in
  Sec.~\ref{sec:shuffle}, all data written for a shuffle
  reader is grouped into one locality set. Then, each shuffle reader
  can sequentially read the data in its corresponding locality set using the
 sequential read service.} 

An example of the user shuffle code has been described in Sec.~\ref{sec:abstraction}. 

\vspace{3pt}
\noindent
\textbf{Virtual Hash Buffer/Hash Service.}
\eat{Partitioned hash aggregation has been widely investigated in the
past~\cite{ graefe2012new}. In these implementations,
the hash table and the
key-value pairs are stored in separate memory areas. For example,
the key-value pairs are written to buffer pool pages, and
the hash table containing only the pointers to keys and values are
allocated from the heap, instead of being allocated from the buffer pool. 
This is problematic, as for many applications, there are going to be a
very large number of distinct keys, and all of this data needs to be paged
efficiently.  Thus, 
it is important to store both the
hash table and key-value pairs on buffer pool pages. 

To address this, }Pangea's hash service adopts
a dynamic partitioning approach,
where each page contains an independent hash table, as well as all of its
associated key-value pairs. We implement this
by using C++ STL unordered-map along with the Memcached slab
allocator~\cite{nishtala2013scaling} to replace the STL default
allocator. The Memcached slab allocator uses the current page as its
memory pool, so all memory allocation is bounded to the memory space
hosting that page. Each page is a hash partition, and all
hash partitions are grouped into one 
locality set. 

We start from $K$ pages as $K$ root partitions, all indexed by a virtual
hash buffer. When there is no free memory in one page, we allocate a
new page
from the buffer pool and split a new child hash partition from the
partition in the page that has used up its memory. We iterate using this process until there is no page that can be allocated
from the buffer pool to construct a new hash partition. Then, when a
page is full, the system needs to select a page, \texttt{unpin} it, and spill it
to disk as partial-aggregation results.

When all objects are inserted through the virtual hash buffer, we 
re-aggregate those spilled partial aggregation results for each
partition. User code is as follows:

\begin{codesmall}
//by below API, "write-back" and "dynamic" 
//allocation policy will be automatically inferred
VirtualHashBufferPtr<string, int> buffer 
  = createVirtualHashBuffer(myOutput);
while((RecordPtr record = myInput.next())) {
  string key = udfGetKey(record);
  int value = udfGetValue(record);
  if( buffer->find(key) == nullptr ) {
    buffer->insert(key, value); 
  } else {
    buffer->set(key, value);
  }
}
\end{codesmall}

\eat{To further reduce the splitting and spilling overhead for large volume
of data, particularly for key-value
pairs that have large but fixed value sizes,  we
store key-value pairs in a separate locality set,
and insert only pointers to
values to the virtual hash buffer. Then when spilling a hash partition, we
spill data directly from the locality set that stores the key-value
pairs to disk and simply free the page holding the hash partition. The
splitting overhead can also be significantly reduced by copying only
pointers. We applied above technique to the $n$-gram counting
application as described in Sec.~\ref{sec:ngram}.}

\eat{
  One or more threads can insert <key, value> pairs
  concurrently to a virtual hash buffer, and as described in
  Sec.~\ref{sec:hash}, the virtual hash buffer
  can automatically grow the hash table by adding new pages and
  splitting partitions that are full in memory. If buffer pool memory becomes
  insufficient, virtual hash buffer will automatically spill
  hash partitions to disk as partial aggregation results.

 After all <key, value> pairs have been inserted, one or more threads
 can read aggregated <key, value> pairs from a virtual hash buffer concurrently
 by allocating hash partitions to those threads. If a partition has
 spills, re-aggregation of partial results is required.}

Pangea also provides other services such as {\it join map service}
for building hash table distributedly from shuffled data;
and {\it broadcast map service}, which broadcasts a locality set and constructs a hash table from it on each node for broadcast join.
Due to space limitation, we omit the details here.

\section{Evaluation}
\label{sec:evaluation}

In this section, we evaluate Pangea. We test
applications such as $k$-means clustering and the TPC-H benchmark
in a distributed cluster, and perform a detailed performance analysis of various
Pangea services in a single node.

For the distributed benchmark, we use $11$ to $31$ AWS r4.2x large
instances, where each instance has eight cores, $61$GB memory, and a $200$GB SSD disk. 
For running
micro-benchmarks of various services,
we use
one AWS m3.xlarge instance that has four CPU cores, $15$GB memory and two
SSD instance store disks. 
On all machines we
install Ubuntu 16.04 and use Spark 2.3.0,
Hadoop 2.7.6, \textcolor{black}{Ignite 2.6.0} and Alluxio 1.7.1.

\subsection{Distributed Benchmark}
\label{sec:distributed}

{\color{black} We have argued in the introduction that a monolithic system such as Pangea should be
considered as an option for building high-performance data analysis tools.  Flexibility may suffer,
but the performance may be excellent.

To investigate whether this claim is reasonable, we implement two computations directly on top of Pangea and compare
the performance 
with a more conventional layered approach: implementing the computation on top of Spark, which is itself using
HDFS or another storage system, all of which is running on top of the JVM.

The first benchmark is a simple $k$-means computation, which is a widely used benchmark for evaluating the 
effect of storage, because one of the main challenges in $k$-means is data locality: keeping as much data
in memory as possible~\cite{zaharia2012resilient}.

For the second benchmark, we actually implement a distributed relational query processor on top of Pangea.  This is particularly interesting because we are implementing a 
reasonably complicated tool on top of Pangea, which makes it possible to quantify the effort of implementing
such a tool as illustrated in Tab.~\ref{tab:sloc}, as well as the performance benefit.}

\vspace{10pt}
\begin{table}[H]
\centering
\small
\caption{\label{tab:sloc} \textcolor{black}{Source Code Break-down for a Pangea-based Relational Query Processor.}}
\begin{tabular}{|p{4cm}|>{\raggedleft\arraybackslash}p{1cm}|} \hline
Component&SLOC\\\hline \hline
Scan& 35\\ \hline
Join&1545\\ \hline
Build broadcast hash map&161 \\\hline
Build partitioned hash map&270 \\\hline
Aggregate: local stage&117\\ \hline
Aggregate: final stage&465 \\ \hline
Filter& 55\\ \hline
Hash& 69\\ \hline
Flatten& 90\\ \hline
Pipeline& 1746\\ \hline
QueryScheduling& 1336\\ \hline
\textbf{Total}&\textbf{5889}\\ \hline
\end{tabular}
\end{table}

\begin{figure}[H]
\centering
\includegraphics[width=3.4in]{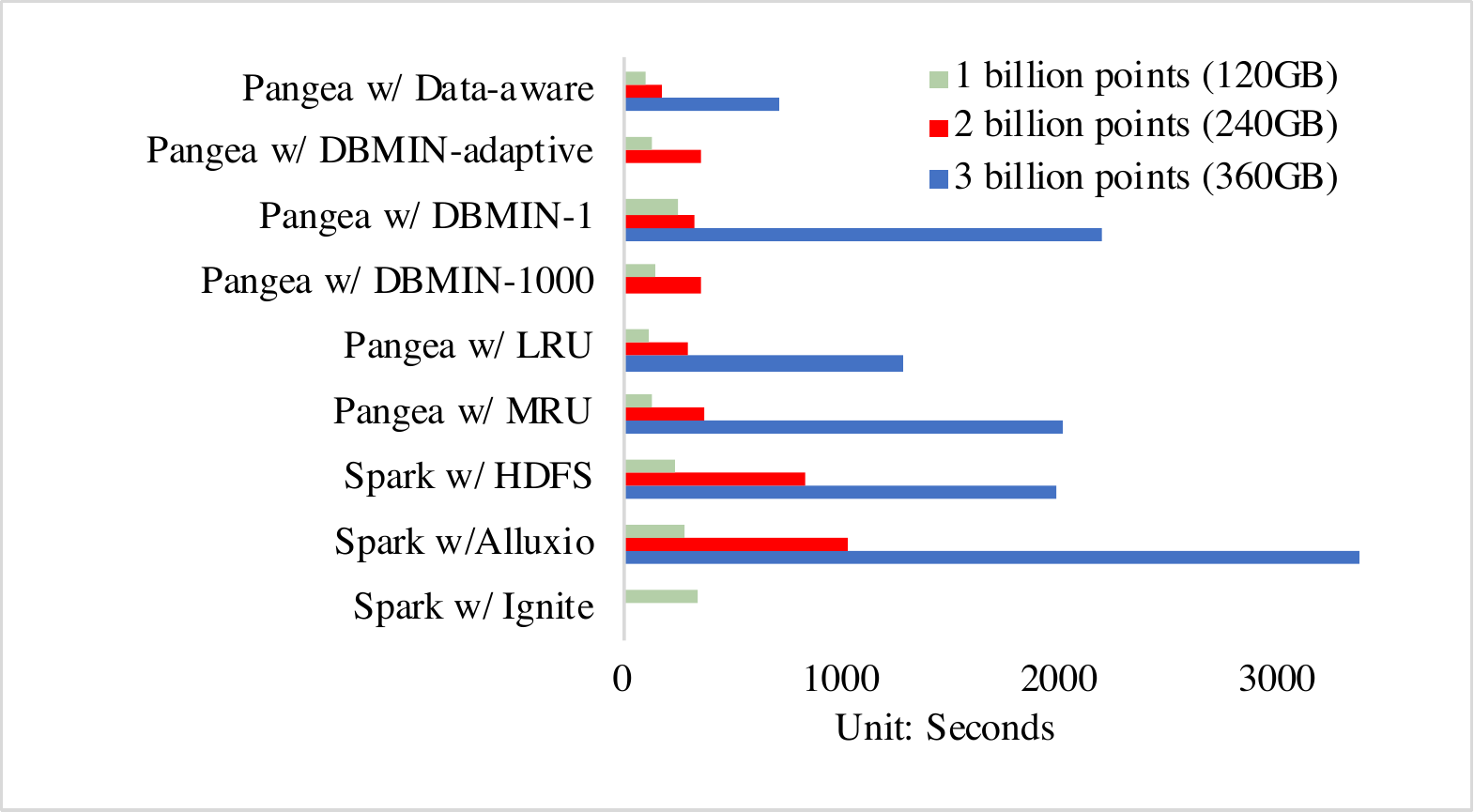}
\caption{Latency comparison for $k$-means with five iterations in 11-node cluster using 1-3 billion 10-dimension points (failed cases are shown as gaps).}
\label{fig:kmeans-all}
\end{figure}

\vspace{3pt}

\subsubsection{$k$-means Clustering}

We develop a $k$-means implementation on Pangea that is similar to the
Spark MLlib implementation. We use double arrays to represent points for
Pangea and wrap a double array in a Hadoop object
as binary input for Spark to minimize (de)serialization
overhead. Each run starts with an initialization
step that computes norms for each point and samples
initial centroids. This is followed by
five iterations of the computation. 

For Pangea, we use one \texttt{Write-through} locality set to store input data, and use one \texttt{Write-back} locality set to store
the points with norms for fast distance computation. For Spark, both
datasets are cached in memory as RDDs.

Spark runs in Yarn client mode and is tested in two different configurations: Spark over HDFS, Spark using
Alluxio as in-memory storage\textcolor{black}{, and Spark using the Ignite SharedRDD}. For each test, we tune Spark
memory allocations for the Spark executor and the OS, Alluxio\textcolor{black}{, or Ignite} for
optimal performance. For
other parameters, we use the default values. Both Spark and Pangea uses $256$MB as split/page size.

As shown in 
Fig.~\ref{fig:kmeans-all}, Pangea facilitates
$k$-means processing at up to a 6$\times$ speedup compared with Spark. It appears that Pangea's monolithic design
facilitates performance gains in several ways:

\vspace{3pt}
\noindent
{\bf (1) Reduction in interfacing overhead}, including overhead for
disk loading, (de)serialization,
  memory (de)allocation, \textcolor{black}{memory compaction for fragmentation,} and
  memory copies. 
In Pangea, user data is directly written to buffer
  pool pages, so when a dataset is imported, a significant portion
  of it is already cached in the buffer pool without any additional
  overhead. In Spark, if an external cache like Alluxio \textcolor{black}{or Ignite} is not being used, data
  cannot be shared across applications, which means user data has to
  be loaded from disk for the initialization step.  We find that
  processing $1$ billion points, the initialization step in Spark over
  HDFS takes $146$ seconds, and each of the following iterations only
  takes $14$ seconds; while in Pangea, for processing the same amount of
  points, the initialization step only takes $43$ seconds, and each
  iteration takes $11$ seconds. This shows how much more efficient Pangea is in moving the data to the application the first time. \textcolor{black}{Based on profiling results, Spark over Ignite spends about 40\% of time in memory compaction due to fragmentation. De-fragmentation occurs because Ignite seems to be 
primarily optimized for frequent random access and updates on mutable data,
and it enforces a 16KB hard page size limitation. In addition, both Spark over Ignite and Alluxio spend significant portion of time in object deserialization.}

\begin{figure}[H]
\centering
\includegraphics[width=3.4in]{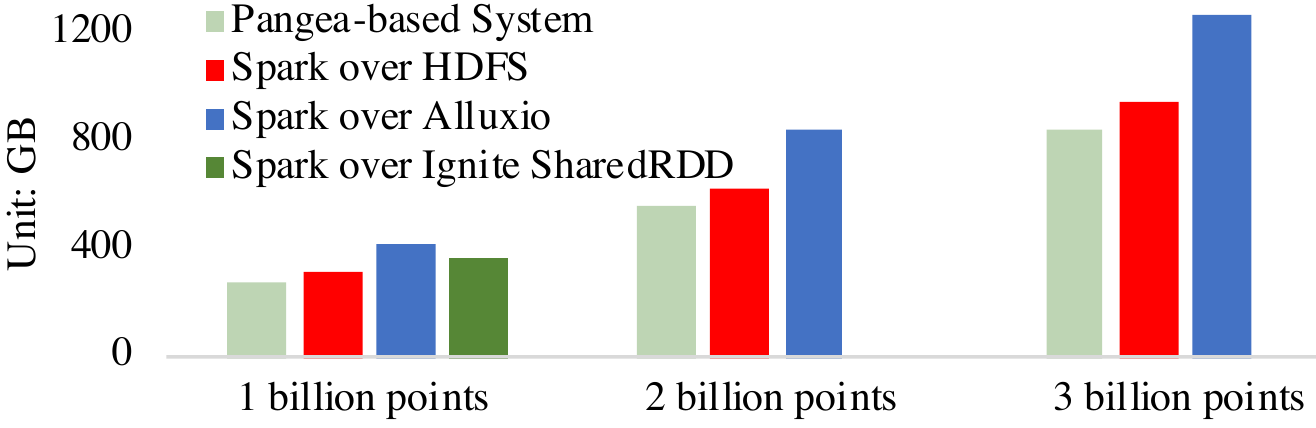}
\caption{Memory usage (failed cases are shown as gaps).}
\label{fig:memory-usage}
\end{figure}
  
\vspace{3pt}
\noindent
{\bf (2) Removal of redundant data placement.}
Although using Alluxio as in-memory storage can avoid data loading
  overhead, double-caching wastes memory resources. Therefore, we
  observe that when processing $1$ billion points using Spark over
  Alluxio, the initialization step time is $96$ seconds, which is
  $1.5 \times$ faster than Spark over HDFS. However, the average
  latency of following iterations is $37$ seconds, which is 3$\times$
  slower than Spark over HDFS due to the fact that Spark has less working memory (we allocated $15$GB memory to
  Alluxio). Fig.~\ref{fig:memory-usage} illustrates the
  memory required by the various setups. The Pangea shared memory pool is configured to be
  $50$GB on each worker node. The total of Spark executor memory
  and Alluxio worker memory is also limited to $50$GB. \textcolor{black}{Ignite requires configuring at least two 
additional memory sizes: the heap size for each Ignite process (we use $5$GB on each node), and the maximum size of the off-heap memory region (we set to $30$GB on each node for one billion points). Ignite throws a segmentation fault when processing 2 billion or more points.}

\vspace{3pt}
\noindent
{\bf (3) Better paging strategy.}
We compare various paging strategies in Fig.~\ref{fig:kmeans-all}. 
\textcolor{black}{We implement three DBMIN algorithms using three size estimation strategies. 
DBMIN-adaptive estimates locality set size exactly following the algorithm in~\cite{chou1986evaluation}, 
while the reference patterns are learned from Pangea-provided services. For DBMIN-1, all 
locality set sizes are estimated as $1$ page. DBMIN-1000 always estimates the size as $1000$ pages. Note that
DBMIN blocks when the total desired size of all locality sets exceeds available size, which is the reason
for the failures of DBMIN-adaptive and DBMIN-1000, as shown in Fig.~\ref{fig:kmeans-all}.}
We find that the data-aware paging
strategy significantly outperforms other paging strategies.
As mentioned in the implementation of $k$-means on both platforms,
input data needs to be first transformed into a new dataset that has
norms associated, which increases the size of working
set. Thus, paging is required at 2 billion points.

\subsubsection{TPC-H}

{\color{black} 
Our Pangea-based relational query processor and API required around 6,000 lines of C++ code to implement eleven different
modules, as illustrated in Tab.~\ref{tab:sloc}. While 6,000 lines may seem like a fairly substantial effort, we have,
in effect, implemented a high-performance distributed query processing engine with that effort.  In addition,
we developed around 
600 lines of shell and python
scripts code for installing and running this computation framework.
 
We then implement nine different TPC-H benchmark queries on top of our analytics engine. Most of those
queries involve aggregation and join.  We compare our implementation with an 
open source, third-party TPC-H implementation\footnote{https://github.com/ssavvides/tpch-spark}, which uses Scala and 
the Spark DataFrame API.}

\begin{figure}[H]
\centering
\includegraphics[width=3.2in]{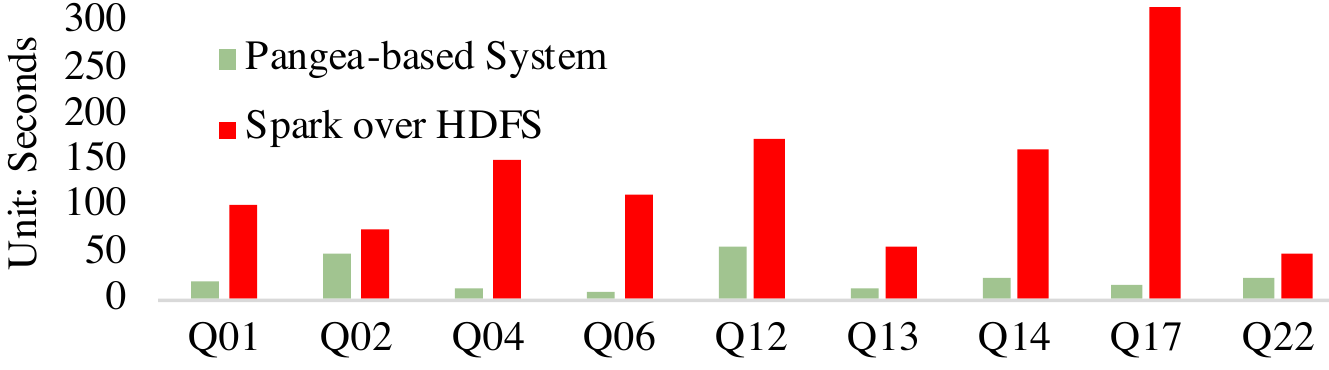}
\caption{Spark vs. Pangea latency (unit: second) for TPC-H queries in
  11-node cluster using scale-100 data.}
\label{fig:tpch-spark-cluster}
\end{figure}

\begin{figure}[H]
\centering
\includegraphics[width=3.2in]{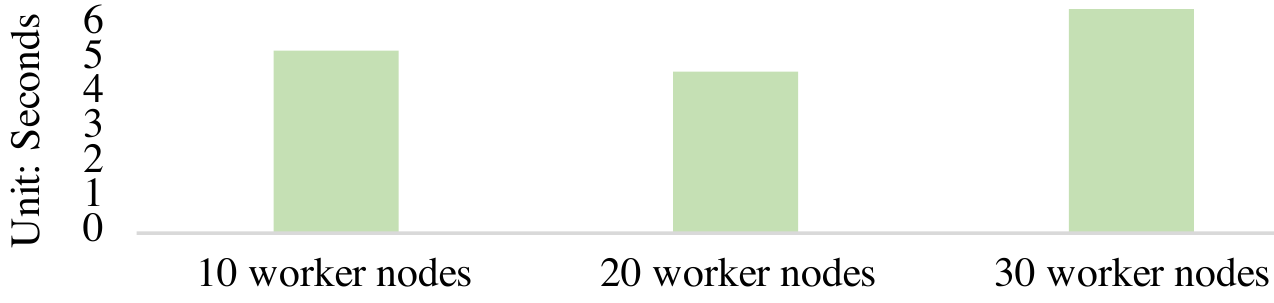}
\caption{Recovery latency (unit: second) for TPC-H scale-100 data in
  clusters with 10 to 30  worker nodes.}
\setlength{\belowcaptionskip}{-10pt}
\label{fig:tpch-recovery}
\end{figure}

In Pangea, the \texttt{lineitem} source set is a randomly dispatched
set which has two replicas, partitioned by
\texttt{l\_orderkey} and \texttt{l\_partkey} respectively; and the \texttt{order} set also has
two replicas, partitioned by \texttt{o\_orderkey} and \texttt{o\_custkey}
respectively. Among the nine TPC-H queries, \texttt{Q04}
and \texttt{Q12} are running on
the \texttt{lineitem} set that is partitioned by \texttt{l\_orderkey}
and the \texttt{order} set that is partitioned by
\texttt{o\_orderkey}; \texttt{Q13} and \texttt{Q22} are running on the \texttt{order} set that is partitioned by
\texttt{o\_custkey}; \texttt{Q14} and \texttt{Q17} are running on the
\texttt{lineitem} set that is partitioned by \texttt{l\_partkey}; and
all other queries run on the source sets.

We generate $100$GB of TPC-H data (Scale-$100$), then compare the
Pangea-based system with Spark over HDFS, and the results are shown
in Fig.~\ref{fig:tpch-spark-cluster}. By using the heterogeneous
replicas for the same table,
Pangea applications can achieve up to $20\times$ speedup compared with
Spark using dataFrame API. Note that there is nothing analogous to
pre-partitioning available to a Spark developer when loading data from
HDFS; all partitioning must be performed at query time. Although a
Spark developer could materialize the repartitioned dataset as a
different HDFS or Alluxio file, the process of loading data into RDD cache is
controlled by Spark task scheduler, which doesn't guarantee locality,
thus repartition at runtime is still needed to perform a local
join. In addition, such manual replica can not be utilized by HDFS for
failure recovery.

Among those queries, for example, Q17 can achieve $20\times$ speedup, mainly
because by selecting the replica of the lineitem set that is
partitioned by \texttt{l\_partkey}, and the replica of the part set
that is partitioned by \texttt{p\_partkey}, the inputs for the
large-scale join in Q17 is co-partitioned, the query scheduler
recognizes this by comparing the available partition schemes of both
sets through the \texttt{statistics} service also provided by Pangea,
and pipelines the join operation at each worker node without
need to do a repartition.

\vspace{5 pt}
\noindent
{\bf Failure Recovery.} As shown in Fig.~\ref{fig:tpch-recovery}, for single-node
failure case, recovering the \texttt{lineitem} table (with 79GB of raw data) in a ten-node cluster using Pangea's
heterogeneous replication only takes five seconds' time, with less than $9\%$ of objects
conflicting. The ratio of conflicting objects declines
significantly with the increase in number of working nodes: $3\%$ for
20 worker nodes, and zero for 30 worker nodes,  as
described in Sec.~\ref{sec:partition}.

These results illustrate that Pangea's heterogeneous replication scheme is
effective. Although using multiple replicas increase storage
size, it is a known and widely accepted cost for high availability.

\subsection{Evaluation of Pangea Services}

In this subsection, we provide some mico-benchmarks of the various Pangea services.

\label{sec:microBenchmarks}
\subsubsection{Sequential Read/Write}
\label{sec:sequential}
This micro-benchmark consists of two tests: one for transient data and one for
persistent data. For both, we first
write a varying number of 80-byte character array objects to different
storage locations, and then we scan
those objects; for each object we compute the sum of all the
bytes. We run the scanning process repeatedly for five times. In the
end, we delete all data.

\begin{figure}[H]
\centering
\includegraphics[width=3.3in]{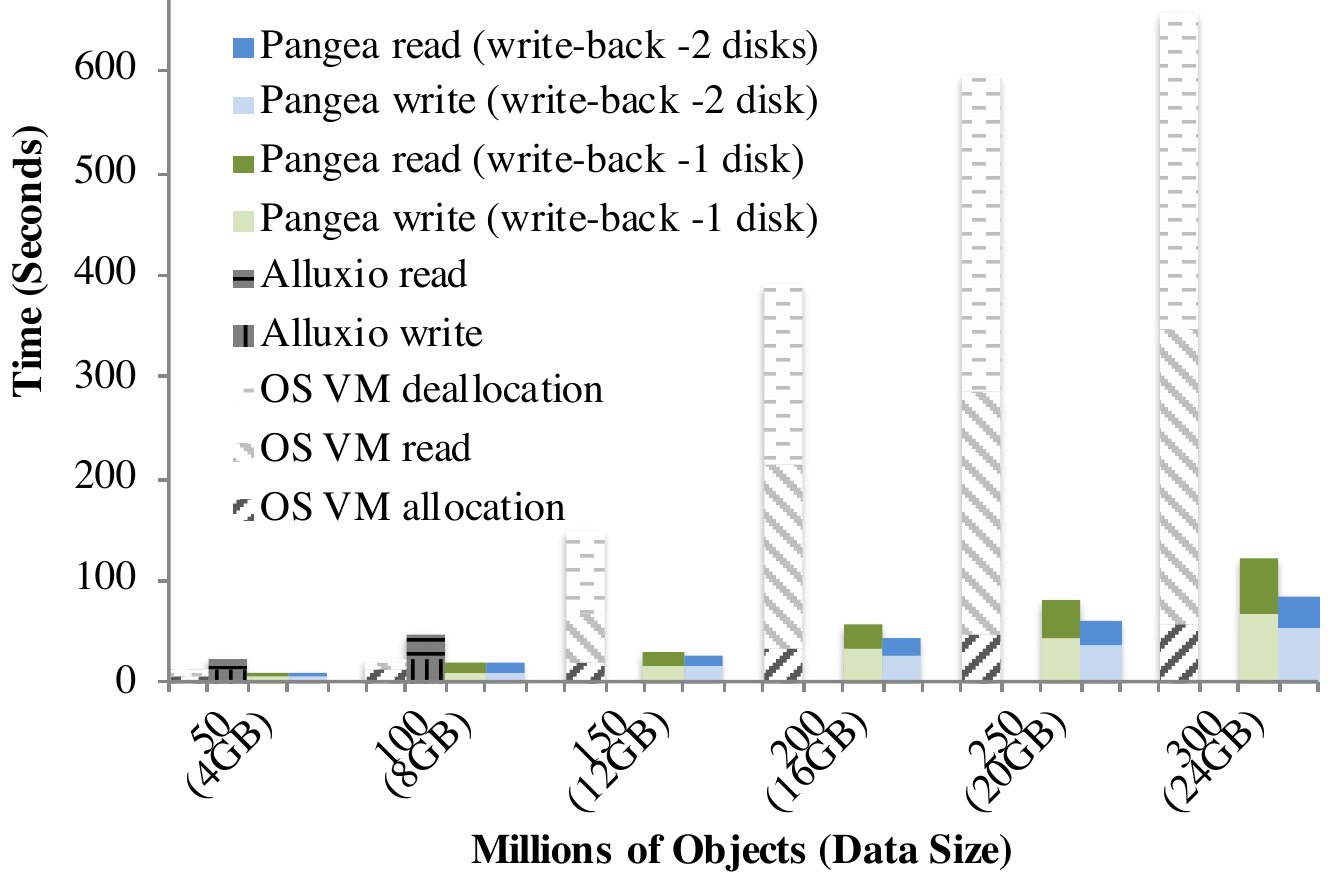}
\caption{Sequential access for
  transient data.}
\label{fig:sequential-transient}
\end{figure}

\vspace{3pt}
\noindent
{\bf Transient Data Test.}

For this test, we
compare Pangea with OS virtual
memory (abbrev. OS VM) and Alluxio. 
For OS VM, we use \texttt{malloc()} and \texttt{free()} for allocation/deallocation.
For Alluxio, we configure the worker memory size to be $14$GB and
develop a Java client that uses a NIO \texttt{ByteBuffer} to efficiently write
data to Alluxio worker, which facilitates a $3\times$ speedup compared with using a JNI-based C++
client.

For Pangea, we use a write-back locality set as the
data container, so that a write request will return immediately once
data is created in the buffer pool.

Results are illustrated in
Fig.~\ref{fig:sequential-transient}.  We observe that when the working set fits in
available memory ($<$ 150 millions of objects in our case), the
performance of Pangea
and OS VM are similar, and both are significantly
better than the Alluxio in-memory file system, presumably because using Alluxio requires
significant interfacing overhead.

When the working set size exceeds
available memory, Pangea can achieve a 5.4$\times$ to
7$\times$
speedup compared with OS VM, mainly because Pangea
increases I/O throughput by using 64MB buffer pool page size and also 
reduces I/O volume through better paging decisions.
Specifically, 
there is only one locality set, for which Pangea automatically
chooses the MRU policy for its sequential access pattern. The OS VM uses LRU policy and other complex techniques such as page
stealing that will evict pages even when there is no
paging demand. In the case of scanning 200 million objects, for each iteration,
Pangea cache will incur 31.4 page-out operations with \textcolor{black}{$2009.6$MB} data
written to disk on
average. However when relying on OS VM, by aggregating the page-out rate collected from the \texttt{sar -B} command contained
in linux \texttt{sysstat} utilities, we see that for each scan iteration,
the average size of data
written to disk by page-out operations is \textcolor{black}{$5074.2$MB} ({2.5$\times$} of Pangea). 

We also observe that using
Pangea 64MB page has a 2.45$\times$
speedup compared with using OS VM 4KB page for 
writing.

Alluxio doesn't support
writing more data than its configured memory size.

Finally, both Pangea and Alluxio are very efficient at removing data.
Because data are organized in large blocks in memory, we can
deallocate data belonging to the
same block at once. This circumvents the cost of individual object
deallocation which accounts for significant overhead, even in C++
applications.

\eat{
\begin{table*}[!htbp]
\centering
\scriptsize
\caption{\label{tab:sequential-rw-memory} Sequential access for
  transient data (unit: seconds) }
\begin{tabular}{|c|c|c|c|c|c|c|c|c|c|c|c|c|c|} \hline
\multirow{2}{*}{NumObjects} &\multicolumn{4}{|c|}{OS virtual memory}&\multicolumn{3}{|c|}{Alluxio}
&\multicolumn{3}{|c|}{Pangea write-back(1
  disk)}&\multicolumn{3}{|c|}{Pangea write-back (2 disks)}\\
\cline {2-14}
& allocation& scan & deallocation&total&write& scan &total&write& scan&total &write& scan &total\\ \hline \hline
50,000,000&7&4&1&{\bf 12}&13&12&{\bf 25}&6&4&{\bf 10}&6&4&10\\ \hline
100,000,000&14&7&3&{\bf 24}&25&23&{\bf 48}&11&8&{\bf 19}&11&8&19 \\ \hline
150,000,000&21&46&83&{\bf 150}&-&-&-&18&11&{\bf 29}&16&11&27\\ \hline
200,000,000&34&181&178&{\bf 417}&-&-&-&33&25&{\bf 58}&27&17&44\\ \hline
250,000,000&46&239&240&{\bf 525}&-&-&-&43&38&{\bf 81}&38&23&61\\ \hline
300,000,000&59&287&311&{\bf 657}&-&-&-&69&52&{\bf 121}&55&30&85\\ \hline
\end{tabular}
\end{table*}
}

\begin{figure}[H]
\centering\subfigure[Writing Persistent Data]{%
   \label{fig:persistent-write}
   \includegraphics[width=3.3in]{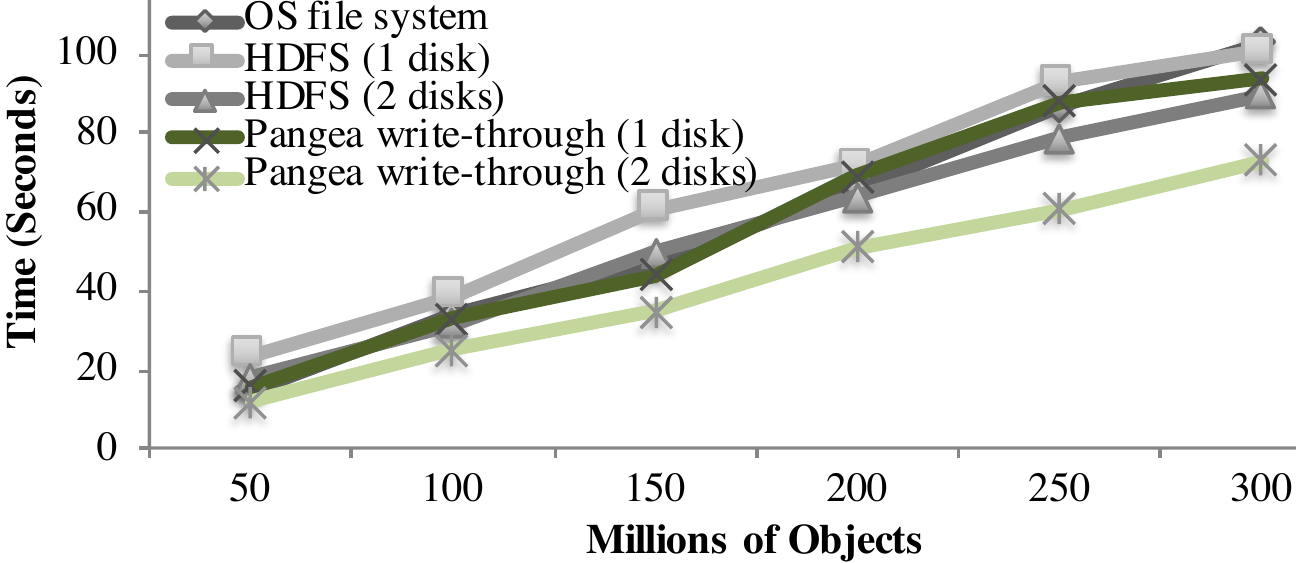} 
}%
\vspace{0pt}
\subfigure[Reading Persistent Data]{%
  \label{fig:persistent-read}
  \includegraphics[width=3.3in]{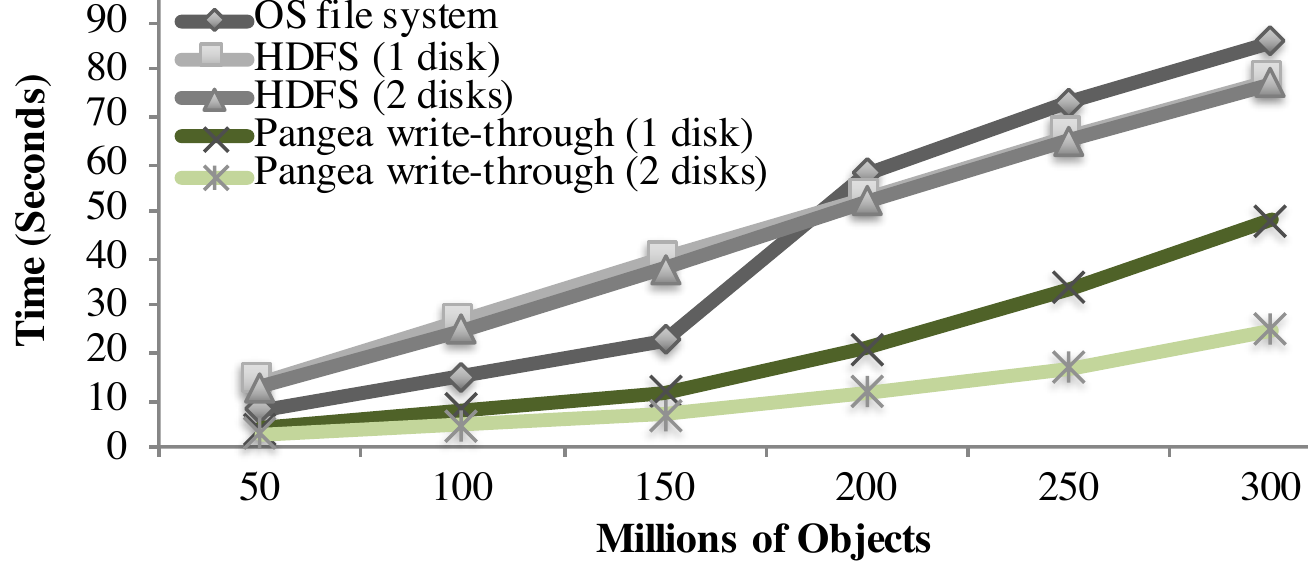}
}
\setlength{\belowcaptionskip}{-10pt}
\caption{\label{fig:sequential-persistent}
Sequential access
  for persistent data
}
\end{figure}

\vspace{3pt}
\noindent
{\bf Persistent Data Test.}
For this test, we use a write-through locality set for Pangea, so that each page will be persisted to
disk via direct I/O immediately when it is fully written. We
compare Pangea with the OS filesystem and HDFS. For HDFS, we use the native C++ client developed
by Cloudera called \texttt{libhdfs3} to avoid JNI overhead and provide a fair comparison.

 The results are illustrated in
Fig.~\ref{fig:sequential-persistent}. After careful tuning, the writing performance of the
three systems is similar. However, for average latency of one
iteration of scan with simple computation,  Pangea outperforms the OS
filesystem 
by a factor of 1.9$\times$ to 2.7$\times$ and outperforms HDFS by 1.5$\times$ to 3.5$\times$.

Through profiling, we find the 
performance gain when the working set fits in memory is mainly from
the reduction
in interfacing overhead; the Pangea
client writes to shared memory directly, and those writes are flushed
to the file system directly.  Thus, we 
can avoid the memory copy overhead between user space and kernel space as
required by the OS buffer cache and also avoid the memory copy between
client and server as required by HDFS. 

When the working set size exceeds available memory size, I/O becomes the
performance bottleneck, and the Pangea Data-aware paging strategy can significantly reduce page
swapping, which is the root cause of the performance gain in
this case.

\eat{
\begin{table*}
\centering
\scriptsize
\caption{\label{tab:sequential-rw-disk} Sequential access
  for persistent data (unit: seconds) }
\begin{tabular}{|c|c|c|c|c|c|c|c|c|c|c|c|c|c|} \hline
\multirow{2}{*}{NumObjects} &\multicolumn{2}{|c|}{OS
  filesystem}&\multicolumn{2}{|c|}{HDFS (1 disk)}&\multicolumn{2}{|c|}{HDFS (2 disks)}
&\multicolumn{2}{|c|}{Pangea write-through(1
  disk)}&\multicolumn{2}{|c|}{Pangea write-through (2 disks)}\\
\cline {2-11}
& write& scan & write& scan &write& scan &write& scan & write& scan \\ \hline \hline
50,000,000&15&{\bf 8}&24&{\bf 14}&18&13&16&{\bf 4}&12&3\\ \hline
100,000,000&34&{\bf 15}&39&{\bf 27}&32&25&33&{\bf 8}&25&5 \\ \hline
150,000,000&47&{\bf 23}&61&{\bf 40}&50&38&44&{\bf 12}&35&7\\ \hline
200,000,000&66&{\bf 58}&72&{\bf 53}&64&52&69&{\bf 21}&51&12\\ \hline
250,000,000&87&{\bf 73}&93&{\bf 66}&79&65&88&{\bf 34}&61&17\\ \hline
300,000,000&103&{\bf 86}&101&{\bf 78}&90&77&94&{\bf 48}&73&25\\ \hline
\end{tabular}
\end{table*}
}

\vspace{3pt}
\noindent
{\bf Paging Strategy Comparison.}
The Data-aware
paging strategy adopts MRU for sequential access pattern. The most
recently used unpinned page will be evicted from a victim locality set
under writing, and at most 10\% of most recently used unpinned pages
will be evicted from a victim locality set under reading. 

We compare above strategy with
MRU, LRU and DBMIN for sequential access. In our implementation,
10\% of most recently used pages will be evicted at each eviction for MRU, and at most 10\%
of least recently used pages will be evicted for LRU. Compared with OS
VM paging, Pangea does not use page stealing, which means it will not evict pages when there is no paging demand.
\textcolor{black}{
There is only one locality set and it is repeatedly read after being written. For such a
loop-sequential pattern, the original DBMIN algorithm suggests configuring the size of the locality set 
to be the set size. For a set exceeding memory size, DBMIN will block.
To avoid this, we upper-bound the locality set size at the memory size.
}

\begin{figure}[H]
\centering\subfigure[Write-through (Persistent)]{%
   \label{fig:sequential-replacement-write-through}
   \includegraphics[width=3.3in]{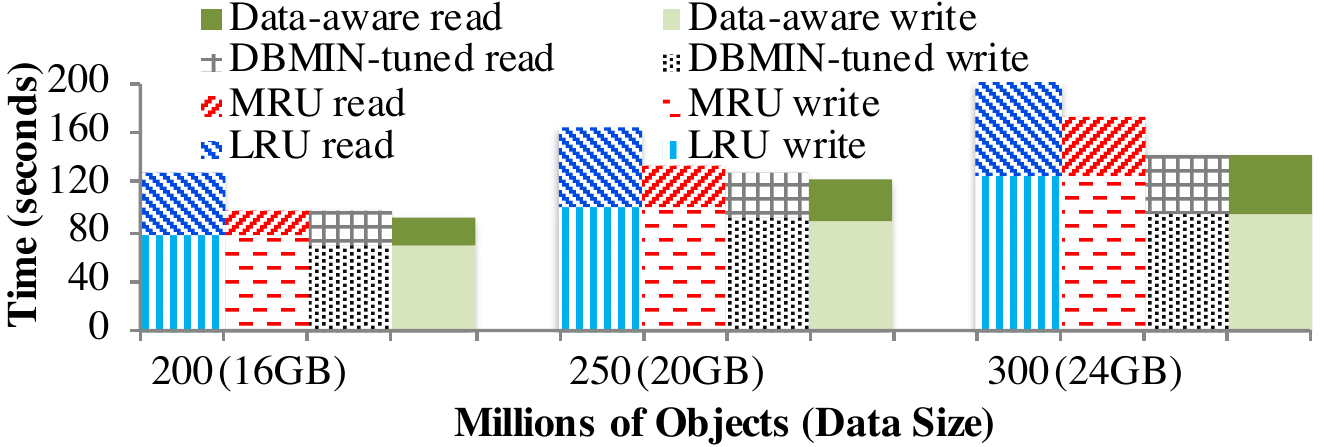}  
}%
\vspace{0pt}
\subfigure[Write-back (Transient)]{%
  \label{fig:sequential-replacement-write-back}
  \includegraphics[width=3.3in]{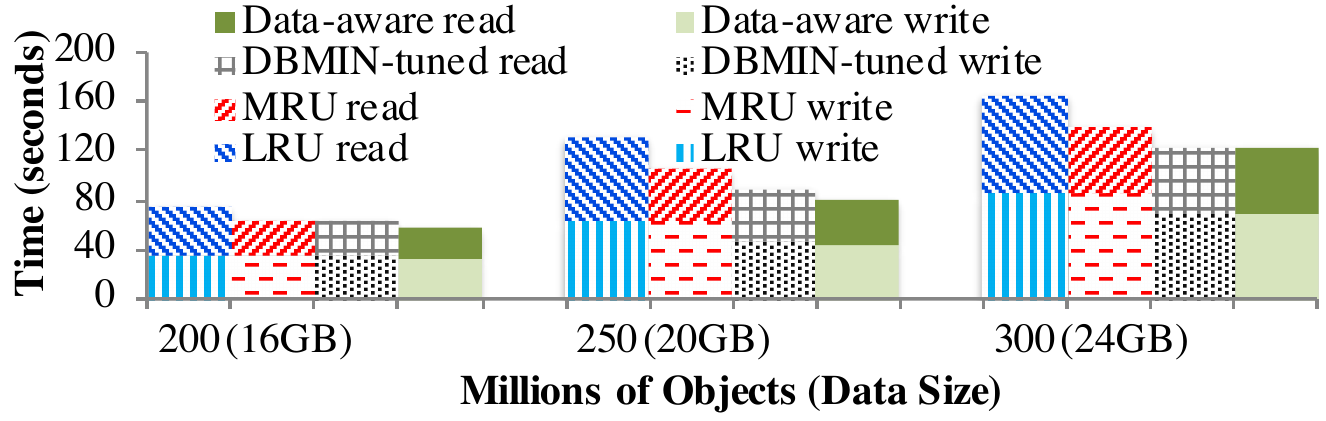}
}
\setlength{\belowcaptionskip}{-10pt}
\caption{\label{fig:sequential-replacement}
  Page replacement for sequential access.
}
\end{figure}

Fig.~\ref{fig:sequential-replacement} lists the comparison results for using one disk. The results for using two disks are similar. For reading, the Pangea
Data-aware policy\textcolor{black}{, tuned DBMIN policy }and MRU can achieve 1.6$\times$ to 2.5$\times$
speedup compared with LRU. This is
because in such a read-after-write scenario---which is common in
dataflow processing---LRU tends to evict pages that will be
read immediately after being evicted.

The Pangea Data-aware
policy can achieve up to $50\%$ performance improvement compared with 
LRU and MRU\textcolor{black}{, and up to $20\%$ compared with the tuned DBMIN algorithm}. This is mainly because with the knowledge about
on-going operations (i.e. read or write), Pangea can
reduce the number of dirty pages to evict to
minimize expensive disk writing operations as compared with LRU and MRU; and can evict more unused pages each time for reading to better overlap I/O operations with computation as compared to DBMIN.

From Fig.~\ref{fig:sequential-replacement}, we see that reading write-back data is slower than
the write-through data. That is because for the former case, transient
pages may be written back to disk in the reading phase, while for the latter case, all
pages are flushed to disk in the writing phase.

\eat{
\begin{table*}
\centering
\small
\caption{\label{tab:sequential-eviction}
 Page replacement comparison for sequential access (unit: seconds) }
\begin{tabular}{|c|c|c|c|c|c|c|c|c|c|c|c|c|} \hline
\multirow{3}{*}{numObjects} &\multicolumn{6}{|c|}{Write-through}
&\multicolumn{6}{|c|}{Write-back}\\
\cline {2-13}
&\multicolumn{2}{|c|}{Pangea LRU}&\multicolumn{2}{|c|}{Pangea
  MRU}&\multicolumn{2}{|c|}{Pangea
  Data-aware}&\multicolumn{2}{|c|}{Pangea
  LRU}&\multicolumn{2}{|c|}{Pangea MRU}&\multicolumn{2}{|c|}{Pangea
  Data-aware}\\
\cline {2-13}
& write& scan & write& scan &write& scan & write&scan & write& scan&write& scan \\ \hline \hline
200,000,000 (16GB)&77&{\bf 52}&77&21&69&{\bf 21}&36&{\bf 39}&36&26&33&{\bf 25} \\ \hline
250,000,000(20GB)&100&{\bf 64}&100&34&88&{\bf 34}&64&{\bf 68}&64&41&43&{\bf 38}\\ \hline
300,000,000 (25GB)&124&{\bf 78}&124&48&94&{\bf 48}&85&{\bf 80}&85&53&69&{\bf 52 }\\ \hline
\end{tabular}
\end{table*}
}
\eat{
{\bf Summary.} Pangaea can provide efficient sequential read and write services
for both transient data (via write-back) and persistent data (via
write-through), when a dataset either fits or exceeds memory. Compared with OS memory management, Pangaea can avoid
deallocation overhead and can achieve up to 7$\times$ speedup in overall
latency. Compared with OS filesystem with buffer caching, Pangaea can
significantly reduce scan latency while providing similar write
performance. Pangaea can also outperform HDFS and Alluxio in most
cases.
}

\subsubsection{Shuffle}
For this micro-benchmark, to provide an apples-to-apples comparison
(not JVM vs. C), we compare Pangea's shuffle service to
simulated Spark shuffling written in C++. In Spark shuffling, each CPU core will have a
separate spill file in the local file system for each shuffle
partition, so there will be $ numCores \times numPartitions$
files in total. For Pangea, all data belonging to the same partition
are written to one locality set, so there are at most $numPartitions$
spill files.

In our test setup, each worker generates small strings of about
10 bytes in length. For each string, a worker computes its partition via a
hash function. For reading shuffle data, each worker reads all strings belonging to one
partition, and for each string, the worker scans each byte and adds up
the byte value.

We use four workers to write to four partitions
and four workers to read from the four partitions. The performance results 
are illustrated in Table.~\ref{tab:shuffle}, which show that we
can achieve {1.1$\times$ - 1.4$\times$} speedup for shuffle writing and {2.2$\times$ - 27$\times$} speedup
for shuffle reading. 

When the working set fits in memory (when the per-thread data size is
smaller than 3500 MB), the performance gain of Pangea is mainly from
the reduction of memory allocation and copy overhead, because in Pangea, the
objects for shuffling are directly allocated in a small page using a
sequential allocator. However, for Spark shuffling, data needs to be
first allocated on heap (we use \texttt{malloc()} for the C++ implementation) and then written to file (we use \texttt{fwrite()}
for the C++ implementation). 

When the working set size exceeds memory (the per-thread data size is larger
than 3500 MB), the performance
gain of Pangea is mainly from significant reduction in I/O overhead that is
brought by a smaller number of files and better page replacement
decisions.

\vspace{5pt}
\begin{table} [H]
\centering
\scriptsize
\caption{\label{tab:shuffle} Shuffle data read and write latency with 4
  writing/reading workers (unit: seconds).}
\begin{tabular}{|r|r|r|r|r|r|r|} \hline
\multirow{2}{*}{MB/thread} &\multicolumn{2}{|c|}{C Spark shuffle} &\multicolumn{2}{|c|}{Pangea (1
  disk)}&\multicolumn{2}{|c|}{Pangea (2 disks)}\\
\cline {2-7}
& write& read & write& read &write& read \\ \hline \hline
500&21&{\bf 5}&15&{\bf $<$1}&15&$<$1\\ \hline
1000&43&{\bf 9}&31&{\bf $<$1}&31&$<$1 \\ \hline
1500&65&{\bf 13}&47&{\bf $<$1}&47&$<$1\\ \hline
2000&86&{\bf 19}&63&{\bf $<$1}&63&$<$1\\ \hline
2500&107&{\bf 23}&80&{\bf $<$1}&79&1\\ \hline
3000&129&{\bf 27}&94&{\bf $<$1}&94&1\\ \hline
3500&152&{\bf 36}&115&{\bf 10}&114&6\\ \hline
4000&172&{\bf 47}&140&{\bf 15}&134&12 \\ \hline
4500&194&{\bf 55}&157&{\bf 20}&155&14\\ \hline
5000&215&{\bf 64}&189&{\bf 27}&174&17\\ \hline
5500&237&{\bf 70}&216&{\bf 32}&196&26\\ \hline
6000&259&{\bf 78}&235&{\bf 35}&215&32\\ \hline
\end{tabular}
\setlength{\belowcaptionskip}{-10pt}
\end{table}

\begin{figure}[H]
\includegraphics[width=3.3in]{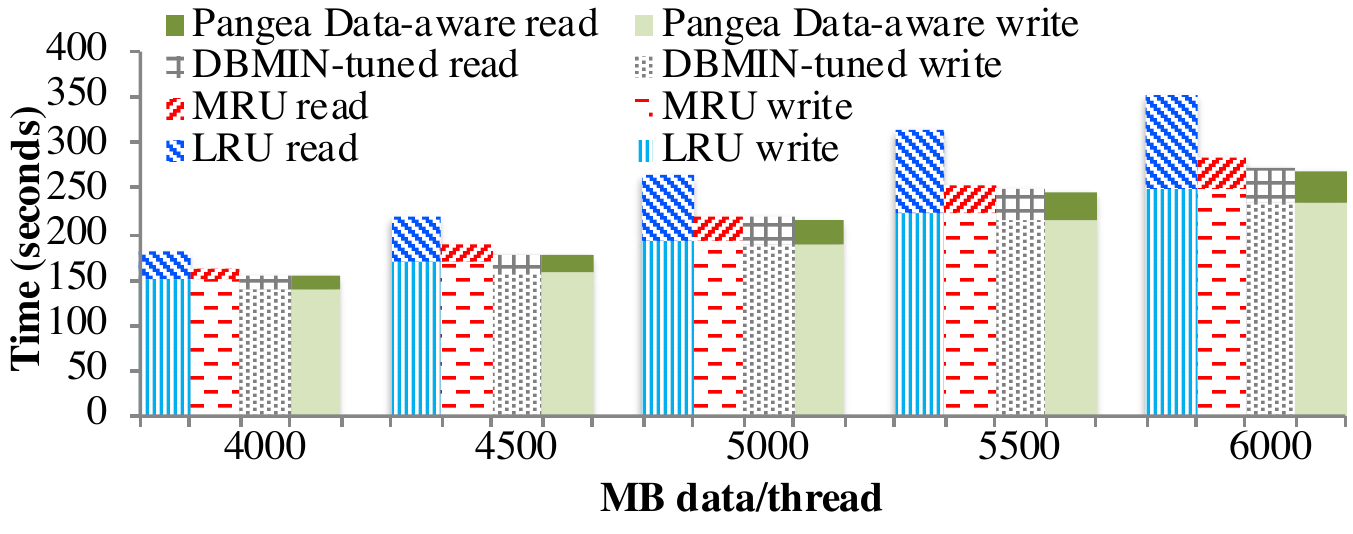}
\caption{\label{fig:shuffle-replacement}
Page replacement comparison for shuffle.
}
\end{figure}

Fig.~\ref{fig:shuffle-replacement} shows the comparison results of different paging policy for shuffle with one disk. The Data-aware
paging policy outperforms LRU in reading
by up to {3$\times$}.
That is because in the reading process, when using a Data-aware policy,
the first 223 pages are kept in
the buffer pool without being flushed and can be directly read by reading
workers. Thus the number of I/O operations is observed to be
significantly reduced. The Pangea Data-aware policy also outperforms
LRU and MRU in writing speed by around $10\%$\textcolor{black}{, and outperforms the tuned DBMIN policy in reading speed by around $10\%$}; that is because by using the former
policy, we can distinguish on-going operations (read or write) through
the locality set's \texttt{CurrentOperation} attribute and can optimize the number of pages to evict for different
operations.

\subsubsection{Hash Aggregation}
We aggregate varying numbers of randomly generated
\texttt{<string, int>} pairs, following the incise.org
benchmark~\cite{HashTableBenchmark}. We compare Pangea with an STL
unordered\_map and Redis 3.2.0, which is a well-known high performance
key-value store developed in C++\cite{sanfilippo2009redis}. The results are illustrated in
Tab.~\ref{tab:hash-insertion}. We see that Pangea hashmap 
outperforms STL unordered map by up to {50$\times$}, and outperforms Redis by
up to {30$\times$}.

The Pangea hashmap is initialized to have
200 partitions. The STL unordered\_map starts to swap
virtual memory when inserting 200 million keys; however the Pangea
hashmap starts spilling to disk only when inserting 300 million
keys. That is mainly because the memcached slab allocator that is used as a 
secondary data allocator in Pangea has better memory utilization
than the STL default allocator. Redis incurs significant latency
because it adopts a client/server architecture, and is inefficient
for problems where the computation can directly run on
local data.

\vspace{5pt}
\begin{table}[H]
\centering
\scriptsize
\caption{\label{tab:hash-insertion} Key-value pair aggregation (unit:
  seconds).}
\begin{tabular}{|r|r|r|r|} \hline
NumKeys & STL unordered\_map& Pangea hashmap& Redis\\\hline \hline
50,000,000&47&{\bf 33}&53\\ \hline
100,000,000&38&{\bf 68}&274 \\ \hline
150,000,000&153&{\bf 110}&2069\\ \hline
200,000,000&7657&{\bf 167}&9103\\ \hline
250,000,000&16818&{\bf 332}&9887\\ \hline
300,000,000&$>$7 hours&{\bf 2450}&failed\\ \hline
\end{tabular}
\setlength{\belowcaptionskip}{-10pt}
\end{table}

\vspace{3pt}
\noindent
\textbf{Summary.}
The experiments show that using Pangea services can
bring up to a $6\times$ speedup for $k$-means, and up to
$20\times$ speedup for TPC-H. In addition, various
micro-benchmarks demonstrate that Pangea can provide high-performance sequential scan, sequential write, shuffle,
hash aggregation services, which are all important building blocks for modern
analytics. 

We also find that in micro-benchmarks that involve only one locality set, Pangea's Data-aware paging policy has similar performance with MRU and DBMIN and is significantly better than LRU. But in distributed benchmarks that involve multiple locality sets each time, Data-aware paging policy can achieve significantly better performance than all of these algorithms in comparison. 

 It may require more effort to build applications on Pangea, but we demonstrate that all changes required by using Pangea services can be
captured in an operator library and end-user is shielded from the changes. This is also proved in
PlinyCompute~\cite{zou2018plinycompute}, which is a separate research work that uses Pangea as
its storage system.


\section{Conclusions}
There are multiple layers in modern data analytics
systems for managing shared persistent data, cached data, and
non-shared execution data in separate systems such as
HDFS, Alluxio\textcolor{black}{, Ignite} and Spark. Such layering introduces
significant performance and management costs for copying data across
layers redundantly. Pangea is designed and implemented to solve this
problem through locality set abstraction at a single layer. Locality set can be
aware of application semantics through services provided within
Pangea, and use this information for
data placement, page
eviction and so on. 

We benchmark Pangea applications including k-means and TPC-H in a distributed cluster with several hundreds of gigabytes data. 
The results show that Pangea is promising to be used as an alternative base for building performance-critical applications and platforms for distributed data analytics.


\balance

\bibliographystyle{abbrv}
\bibliography{refs}  

\begin{thebibliography}{10}

\bibitem{s3}
Amazon simple storage system.
\newblock https://aws.amazon.com/s3.

\bibitem{ignite}
Apache ignite.
\newblock https://ignite.apache.org.

\bibitem{GoogleCouldStorage}
Google cloud storage.
\newblock https://cloud.google.com/storage.

\bibitem{HashTableBenchmark}
Hash table benchmark.
\newblock http://incise.org/hash-table-benchmarks.html.

\bibitem{tungsten}
Project tungsten: Bringing spark closer to bare metal.
\newblock
  https://databricks.com/blog/2015/04/28/project-tungsten-bringing-spark-closer-to-bare-metal.html.

\bibitem{clusterSize}
Why enterprises of different sizes are adopting 'fast data' with apache spark.
\newblock
  https://www.lightbend.com/blog/why-enterprises-of-different-sizes-are-adopting-fast-data-with-apache-spark.

\bibitem{abaditensorflow}
M.~Abadi, A.~Agarwal, P.~Barham, E.~Brevdo, Z.~Chen, C.~Citro, G.~S. Corrado,
  A.~Davis, J.~Dean, M.~Devin, et~al.
\newblock Tensorflow: Large-scale machine learning on heterogeneous systems,
  2015.
\newblock {\em Software available from tensorflow. org}.

\bibitem{agrawal2004integrating}
S.~Agrawal, V.~Narasayya, and B.~Yang.
\newblock Integrating vertical and horizontal partitioning into automated
  physical database design.
\newblock In {\em Proceedings of the 2004 ACM SIGMOD international conference
  on Management of data}, pages 359--370. ACM, 2004.

\bibitem{alexandrov2014stratosphere}
A.~Alexandrov and et~al.
\newblock The stratosphere platform for big data analytics.
\newblock {\em VLDB}, 23(6):939--964, 2014.

\bibitem{ananthanarayanan2012pacman}
G.~Ananthanarayanan, A.~Ghodsi, A.~Wang, D.~Borthakur, S.~Kandula, S.~Shenker,
  and I.~Stoica.
\newblock Pacman: Coordinated memory caching for parallel jobs.
\newblock In {\em Proceedings of the 9th USENIX conference on Networked Systems
  Design and Implementation}, pages 20--20. USENIX Association, 2012.

\bibitem{armbrust2015spark}
M.~Armbrust, R.~S. Xin, C.~Lian, Y.~Huai, D.~Liu, J.~K. Bradley, X.~Meng,
  T.~Kaftan, M.~J. Franklin, A.~Ghodsi, et~al.
\newblock Spark sql: Relational data processing in spark.
\newblock In {\em Proceedings of the 2015 ACM SIGMOD International Conference
  on Management of Data}, pages 1383--1394. ACM, 2015.

\bibitem{arnold2014openstack}
J.~Arnold.
\newblock {\em Openstack swift: Using, administering, and developing for swift
  object storage}.
\newblock " O'Reilly Media, Inc.", 2014.

\bibitem{bent2004explicit}
J.~Bent, D.~Thain, A.~C. Arpaci-Dusseau, R.~H. Arpaci-Dusseau, and M.~Livny.
\newblock Explicit control in the batch-aware distributed file system.
\newblock In {\em NSDI}, volume~4, pages 365--378, 2004.

\bibitem{borthakur2008hdfs}
D.~Borthakur.
\newblock Hdfs architecture guide.
\newblock {\em HADOOP APACHE PROJECT http://hadoop. apache.
  org/common/docs/current/hdfs design. pdf}, 2008.

\bibitem{bovet2005understanding}
D.~P. Bovet and M.~Cesati.
\newblock {\em Understanding the Linux kernel}.
\newblock " O'Reilly Media, Inc.", 2005.

\bibitem{calder2011windows}
B.~Calder, J.~Wang, A.~Ogus, N.~Nilakantan, A.~Skjolsvold, S.~McKelvie, Y.~Xu,
  S.~Srivastav, J.~Wu, H.~Simitci, et~al.
\newblock Windows azure storage: a highly available cloud storage service with
  strong consistency.
\newblock In {\em Proceedings of the Twenty-Third ACM Symposium on Operating
  Systems Principles}, pages 143--157. ACM, 2011.

\bibitem{cao1996implementation}
P.~Cao and et~al.
\newblock Implementation and performance of integrated application-controlled
  file caching, prefetching, and disk scheduling.
\newblock {\em TOCS}, 14(4):311--343, 1996.

\bibitem{cao1997cost}
P.~Cao and S.~Irani.
\newblock Cost-aware www proxy caching algorithms.
\newblock In {\em Usenix symposium on internet technologies and systems},
  volume~12, pages 193--206, 1997.

\bibitem{chaiken2008scope}
R.~Chaiken, B.~Jenkins, P.-{\AA}. Larson, B.~Ramsey, D.~Shakib, S.~Weaver, and
  J.~Zhou.
\newblock Scope: easy and efficient parallel processing of massive data sets.
\newblock {\em Proceedings of the VLDB Endowment}, 1(2):1265--1276, 2008.

\bibitem{chen2012interactive}
Y.~Chen and et~al.
\newblock Interactive analytical processing in big data systems: A
  cross-industry study of mapreduce workloads.
\newblock {\em VLDB}, 5(12):1802--1813, 2012.

\bibitem{chou1986evaluation}
H.-T. Chou and D.~J. DeWitt.
\newblock An evaluation of buffer management strategies for relational database
  systems.
\newblock {\em Algorithmica}, 1(1-4):311--336, 1986.

\bibitem{crotty2015architecture}
A.~Crotty, A.~Galakatos, K.~Dursun, T.~Kraska, C.~Binnig, U.~Cetintemel, and
  S.~Zdonik.
\newblock An architecture for compiling udf-centric workflows.
\newblock {\em Proceedings of the VLDB Endowment}, 8(12):1466--1477, 2015.

\bibitem{dittrich2010hadoop++}
J.~Dittrich, J.-A. Quian{\'e}-Ruiz, A.~Jindal, Y.~Kargin, V.~Setty, and
  J.~Schad.
\newblock Hadoop++: making a yellow elephant run like a cheetah (without it
  even noticing).
\newblock {\em Proceedings of the VLDB Endowment}, 3(1-2):515--529, 2010.

\bibitem{ellard2003attribute}
D.~Ellard, E.~Thereska, G.~R. Ganger, M.~I. Seltzer, et~al.
\newblock Attribute-based prediction of file properties.
\newblock 2003.

\bibitem{eltabakh2011cohadoop}
M.~Y. Eltabakh, Y.~Tian, F.~{\"O}zcan, R.~Gemulla, A.~Krettek, and
  J.~McPherson.
\newblock Cohadoop: flexible data placement and its exploitation in hadoop.
\newblock {\em Proceedings of the VLDB Endowment}, 4(9):575--585, 2011.

\bibitem{fagin1978efficient}
R.~Fagin and T.~G. Price.
\newblock Efficient calculation of expected miss ratios in the independent
  reference model.
\newblock {\em SIAM Journal on Computing}, 7(3):288--297, 1978.

\bibitem{fitzpatrick2004distributed}
B.~Fitzpatrick.
\newblock Distributed caching with memcached.
\newblock {\em Linux journal}, 2004(124):5, 2004.

\bibitem{fonseca2002intrinsic}
R.~Fonseca, V.~Almeida, M.~Crovella, and B.~Abrahao.
\newblock On the intrinsic locality properties of web reference streams.
\newblock Technical report, Boston University Computer Science Department,
  2002.

\bibitem{garetto2015efficient}
M.~Garetto, E.~Leonardi, and S.~Traverso.
\newblock Efficient analysis of caching strategies under dynamic content
  popularity.
\newblock In {\em Computer Communications (INFOCOM), 2015 IEEE Conference on},
  pages 2263--2271. IEEE, 2015.

\bibitem{ghemawat2003google}
S.~Ghemawat and et~al.
\newblock The google file system.
\newblock In {\em ACM SIGOPS Operating Systems Review}, volume~37, pages
  29--43. ACM, 2003.

\bibitem{gupta2011gpfs}
K.~Gupta and et~al.
\newblock Gpfs-snc: An enterprise storage framework for virtual-machine clouds.
\newblock {\em IBM Journal of Research and Development}, 55(6):2--1, 2011.

\bibitem{jaleel2010high}
A.~Jaleel, K.~B. Theobald, S.~C. Steely~Jr, and J.~Emer.
\newblock High performance cache replacement using re-reference interval
  prediction (rrip).
\newblock In {\em ACM SIGARCH Computer Architecture News}, volume~38, pages
  60--71. ACM, 2010.

\bibitem{jindal2018computation}
A.~Jindal, S.~Qiao, H.~Patel, Z.~Yin, J.~Di, M.~Bag, M.~Friedman, Y.~Lin,
  K.~Karanasos, and S.~Rao.
\newblock Computation reuse in analytics job service at microsoft.
\newblock In {\em Proceedings of the 2018 International Conference on
  Management of Data}, pages 191--203. ACM, 2018.

\bibitem{jyothi2016morpheus}
S.~A. Jyothi, C.~Curino, I.~Menache, S.~M. Narayanamurthy, A.~Tumanov,
  J.~Yaniv, R.~Mavlyutov, I.~Goiri, S.~Krishnan, J.~Kulkarni, et~al.
\newblock Morpheus: Towards automated slos for enterprise clusters.
\newblock In {\em OSDI}, pages 117--134, 2016.

\bibitem{kleinrock1976queueing}
L.~Kleinrock.
\newblock {\em Queueing systems, volume 2: Computer applications}, volume~66.
\newblock Wiley New York, 1976.

\bibitem{kornacker2012cloudera}
M.~Kornacker and J.~Erickson.
\newblock Cloudera impala: Real time queries in apache hadoop, for real.
\newblock {\em ht tp://blog. cloudera.
  com/blog/2012/10/cloudera-impala-real-time-queries-in-apache-hadoop-for-real},
  2012.

\bibitem{lee2001lrfu}
D.~Lee, J.~Choi, J.-H. Kim, S.~H. Noh, S.~L. Min, Y.~Cho, and C.~S. Kim.
\newblock Lrfu: A spectrum of policies that subsumes the least recently used
  and least frequently used policies.
\newblock {\em IEEE transactions on Computers}, (12):1352--1361, 2001.

\bibitem{li2018alluxio}
H.~Li.
\newblock Alluxio: A virtual distributed file system.
\newblock 2018.

\bibitem{li2014tachyon}
H.~Li and et~al.
\newblock Tachyon: Reliable, memory speed storage for cluster computing
  frameworks.
\newblock In {\em SOCC}, pages 1--15, 2014.

\bibitem{liedtke1996toward}
J.~Liedtke.
\newblock Toward real microkernels.
\newblock {\em Communications of the ACM}, 39(9):70--77, 1996.

\bibitem{lu2016lifetime}
L.~Lu and et~al.
\newblock Lifetime-based memory management for distributed data processing
  systems.
\newblock {\em VLDB}, 9(12):936--947, 2016.

\bibitem{masmano2004tlsf}
M.~Masmano, I.~Ripoll, A.~Crespo, and J.~Real.
\newblock Tlsf: A new dynamic memory allocator for real-time systems.
\newblock In {\em Real-Time Systems, 2004. ECRTS 2004. Proceedings. 16th
  Euromicro Conference on}, pages 79--88. IEEE, 2004.

\bibitem{mesnier2004file}
M.~Mesnier, E.~Thereska, G.~R. Ganger, D.~Ellard, and M.~Seltzer.
\newblock File classification in self-* storage systems.
\newblock In {\em Autonomic Computing, 2004. Proceedings. International
  Conference on}, pages 44--51. IEEE, 2004.

\bibitem{mortonusermode}
A.~Morton.
\newblock Usermode pagecache control: fadvise ().

\bibitem{nishtala2013scaling}
R.~Nishtala and et~al.
\newblock Scaling memcache at facebook.
\newblock In {\em NSDI}, pages 385--398, 2013.

\bibitem{o1993lru}
E.~J. O'neil and et~al.
\newblock The lru-k page replacement algorithm for database disk buffering.
\newblock {\em ACM SIGMOD Record}, 22(2):297--306, 1993.

\bibitem{pai2000io}
V.~S. Pai, P.~Druschel, and W.~Zwaenepoel.
\newblock Io-lite: a unified i/o buffering and caching system.
\newblock {\em ACM Transactions on Computer Systems (TOCS)}, 18(1):37--66,
  2000.

\bibitem{pedregosa2011scikit}
F.~Pedregosa, G.~Varoquaux, A.~Gramfort, V.~Michel, B.~Thirion, O.~Grisel,
  M.~Blondel, P.~Prettenhofer, R.~Weiss, V.~Dubourg, et~al.
\newblock Scikit-learn: Machine learning in python.
\newblock {\em Journal of machine learning research}, 12(Oct):2825--2830, 2011.

\bibitem{rao2002automating}
J.~Rao, C.~Zhang, N.~Megiddo, and G.~Lohman.
\newblock Automating physical database design in a parallel database.
\newblock In {\em Proceedings of the 2002 ACM SIGMOD international conference
  on Management of data}, pages 558--569. ACM, 2002.

\bibitem{sanfilippo2009redis}
S.~Sanfilippo and P.~Noordhuis.
\newblock Redis, 2009.

\bibitem{sathiamoorthy2013xoring}
M.~Sathiamoorthy, M.~Asteris, D.~Papailiopoulos, A.~G. Dimakis, R.~Vadali,
  S.~Chen, and D.~Borthakur.
\newblock Xoring elephants: Novel erasure codes for big data.
\newblock In {\em Proceedings of the VLDB Endowment}, volume~6, pages 325--336.
  VLDB Endowment, 2013.

\bibitem{stonebraker2005c}
M.~Stonebraker, D.~J. Abadi, A.~Batkin, X.~Chen, M.~Cherniack, M.~Ferreira,
  E.~Lau, A.~Lin, S.~Madden, E.~O'Neil, et~al.
\newblock C-store: a column-oriented dbms.
\newblock In {\em Proceedings of the 31st international conference on Very
  large data bases}, pages 553--564. VLDB Endowment, 2005.

\bibitem{weil2006ceph}
S.~A. Weil, S.~A. Brandt, E.~L. Miller, D.~D. Long, and C.~Maltzahn.
\newblock Ceph: A scalable, high-performance distributed file system.
\newblock In {\em Proceedings of the 7th symposium on Operating systems design
  and implementation}, pages 307--320. USENIX Association, 2006.

\bibitem{white2012hadoop}
T.~White.
\newblock {\em {Hadoop: The Definitive Guide}}.
\newblock O'Reilly Media, 2012.

\bibitem{wu2013studying}
M.-J. Wu, M.~Zhao, and D.~Yeung.
\newblock Studying multicore processor scaling via reuse distance analysis.
\newblock In {\em ACM SIGARCH Computer Architecture News}, volume~41, pages
  499--510. ACM, 2013.

\bibitem{young1994thek}
N.~Young.
\newblock The k-server dual and loose competitiveness for paging.
\newblock {\em Algorithmica}, 11(6):525--541, 1994.

\bibitem{zaharia2012resilient}
M.~Zaharia and et~al.
\newblock Resilient distributed datasets: A fault-tolerant abstraction for
  in-memory cluster computing.
\newblock In {\em NSDI}, pages 2--15. USENIX, 2012.

\bibitem{zhou2012advanced}
J.~Zhou, N.~Bruno, and W.~Lin.
\newblock Advanced partitioning techniques for massively distributed
  computation.
\newblock In {\em Proceedings of the 2012 ACM SIGMOD International Conference
  on Management of Data}, pages 13--24. ACM, 2012.

\bibitem{zhou2001multi}
Y.~Zhou, J.~Philbin, and K.~Li.
\newblock The multi-queue replacement algorithm for second level buffer caches.
\newblock In {\em USENIX Annual Technical Conference, General Track}, pages
  91--104, 2001.

\bibitem{zou2018plinycompute}
J.~Zou, R.~M. Barnett, T.~Lorido-Botran, S.~Luo, C.~Monroy, S.~Sikdar,
  K.~Teymourian, B.~Yuan, and C.~Jermaine.
\newblock Plinycompute: A platform for high-performance, distributed,
  data-intensive tool development.
\newblock In {\em Proceedings of the 2018 International Conference on
  Management of Data}, pages 1189--1204. ACM, 2018.

\end{thebibliography}

\end{document}